\documentclass[aps,pre,twocolumn,showkeys,showpacs,a4paper,floatfix,amsmath,amssymb,superscriptaddress]{revtex4}
\usepackage{natbib}
\usepackage{dcolumn}
\usepackage{graphicx}

\begin{document}

\title{Efficient data processing and quantum phenomena:\\
Single-particle systems}

\author{H. De Raedt}
\email{h.a.de.raedt@rug.nl}
\homepage{http://www.compphys.org}
\affiliation{Department of Applied Physics,
Materials Science Centre, University of Groningen, Nijenborgh 4,
NL-9747 AG Groningen, The Netherlands}

\author{K. De Raedt}
\email{deraedt@cs.rug.nl}
\affiliation{Department of Computer Science,
University of Groningen, Blauwborgje 3,
NL-9747 AC Groningen, The Netherlands}

\author{K. Michielsen}
\email{k.f.l.michielsen@rug.nl}
\affiliation{Department of Applied Physics,
Materials Science Centre, University of Groningen, Nijenborgh 4,
NL-9747 AG Groningen, The Netherlands}

\author{S. Miyashita}
 \email{miya@spin.phys.t.u-tokyo.ac.jp}
\affiliation{Department of Physics, Graduate School of Science,
University of Tokyo, Bunkyo-ku Tokyo 113-8656, Japan}

\begin{abstract}
We study the relation between the acquisition and analysis of data
and quantum theory using a probabilistic and deterministic model
for photon polarizers.
We introduce criteria for efficient processing of data and then
use these criteria to demonstrate that efficient processing
of the data contained in single events is equivalent to
the observation that Malus' law holds.
A strictly deterministic process that also yields Malus' law
is analyzed in detail.
We present a performance analysis of the
probabilistic and deterministic model of the photon polarizer.
The latter is an adaptive dynamical system
that has primitive learning capabilities.
This additional feature has recently been shown to be
sufficient to perform event-by-event simulations of
interference phenomena, without using concepts of wave mechanics.
We illustrate this by presenting results for a system of
two chained Mach-Zehnder interferometers, suggesting
that systems that perform efficient data processing and
have learning capability are able to exhibit behavior
that is usually attributed to quantum systems only.

\end{abstract}
\date{\today}

\maketitle

\def\ORDER#1{\hbox{${\cal O}(#1)$}}
\def\BRA#1{\langle #1 \vert}
\def\KET#1{\vert #1 \rangle}
\def\EXPECT#1{\langle #1 \rangle}
\def\BRACKET#1#2{\langle #1 \vert #2 \rangle}
\def\hbar{{\mathchar'26\mskip-9muh}}
\def\mod{{\mathop{\hbox{mod}}}}
\def\CNOT{{\mathop{\hbox{CNOT}}}}
\def\Tr{{\mathop{\hbox{Tr}}}}
\def\bPsi{{\mathbf{\Psi}}}
\def\bPhi{{\mathbf{\Phi}}}
\def\bzero{{\mathbf{0}}}
\def\Eq#1{(\ref{#1})}
\def\NOBAR#1{#1}
\def\BAR#1{\overline{#1}}
\def\openone{\leavevmode\hbox{\small1\kern-3.8pt\normalsize1}}
\def\DLM{DLM}
\def\DLMS{DLMs}

\section{Introduction}\label{sec1} 

\begin{figure}[t]
\begin{center}
\includegraphics[width=8cm]{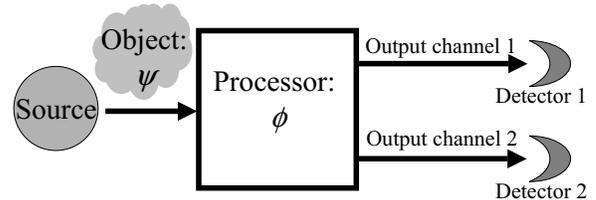}
\caption{
Schematic representation of the event-by-event experiment.
The source emits objects one at a time.
Each object carries a message
represented by an angle $\psi$.
The processor combines this message
with the angle $\phi$ that is controlled by the user,
and sends the object through one of its two
output channels.
Detectors count the number of objects
in each channel.
}
\label{machine0}
\end{center}
\end{figure}

Consider the schematic representation of an experiment
in which a source emits objects that carry information
represented by an angle $0\le\psi\le2\pi$ (see Fig.~\ref{machine0}).
We want to determine this angle as accurately as possible,
but there are some limitations on the equipment that
is available to us, namely:

\setlength{\unitlength}{1cm}

\begin{itemize}
\item We do not have a device that can measure the angle $\psi$ directly.
\item We have detectors that can count the arrival (or passage)
of individual objects.
\item We can build a device, called processor in what follows,
that can direct an incoming object
to one of its two output channels according to $\psi$,
relative to the orientation $0\le\phi\le2\pi$ of the processor itself.
We can count the number of objects in each output channel by
using the detectors.
\item We do not have prior information about the angle $\psi$ itself,
implying that there is no reason to assume that
particular angles $\psi$ are more likely to occur than others.
Note that this does not imply that $\psi$ is a random variable.
\end{itemize}

Given this scenario, the obvious question is:
What kind of functionality should we build into the processor
such that by counting $N$ events in the output channels,
we obtain an accurate estimate for the angle $\psi$ ?
Of course, the optimal design of the processor
depends on additional constraints.
Usually, we prefer devices of which the output does
not change drastically when the input varies a little.
Therefore we require that
\begin{enumerate}
\item The number of events generated in each output channel
is most insensitive to small changes in $\theta\equiv\psi-\phi$.
\item The performance of the processor should be
insensitive to the actual value of $\theta$.
\end{enumerate}
Using these two criteria, we consider two extreme realizations of the processor.
First, we construct a simple probabilistic processor
that operates according to the rules of probability theory
and uses random numbers to transform the input data $\psi$ into
a sequence of discrete output events.
Then, we present a strictly deterministic processor that performs
the same task as the simple probabilistic processor.
In both cases, the general strategy to determine the optimal
processor is the same: We search for a probabilistic or deterministic process
that satisfies the criteria (1) and (2) mentioned earlier.
Then we estimate the efficiency of the processor.
As convenient measure for the efficiency (or performance)
of a processor, we take the number of different messages $M_D$ that can be
extracted from a record of $N$ bits, each bit representing an event
in one of the two output channels, with a specified level of certainty.

The reader may have noticed that the scenario
we described earlier applies to the
measurement of the polarization of light
in the regime where the signal from the detectors consists of
discrete ``clicks''~\cite{BAYM74,FEYN65}.
In this case, the objects are represented by photons,
the angle $\psi$ describes the polarization
(which we cannot measure directly),
the detectors may be photon multiplier tubes or semiconductor diodes,
and the processor a properly prepared calcite crystal~\cite{BORN64}.
In general terms, we want to characterize the behavior of a system in terms of
numerical quantities that we can obtain
by repeating measurements that give us partial information only.
This is a characteristic feature of quantum mechanics.
In order not to become entangled in the difficulties
with the interpretation of quantum theory and
the measurement paradox in particular~\cite{HOME97},
in the theoretical analysis presented in this paper,
we avoid the use of words such as photons and polarization.
A remarkable result of this paper is
that the search for an efficient (in the sense specified earlier),
data processor yields probabilistic and deterministic processors
that generate output events according to Malus' law.

\section{Probabilistic processor}{\label{sec2}}

\begin{figure}[t]
\begin{center}
\includegraphics[width=9cm]{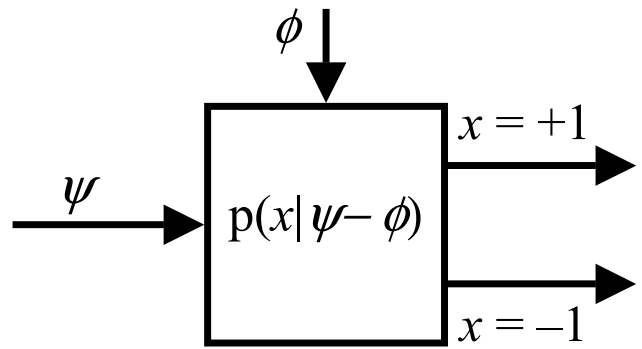}
\caption{
Schematic diagram of a probabilistic processor that transforms
the difference between the input angle $\psi$ and
the setting $\phi$ into a sequence of $x=\pm1$ signals.
}
\label{machine1}
\end{center}
\end{figure}

The schematic diagram of the probabilistic processor is shown
in Fig.~\ref{machine1}.
The input to the processor is an event that
carries a message represented by the angle $\psi$.
The presence of an event in one of the output channels
is represented by a message that carries the variable $x=\pm1$.
We assume that the experimental data is in concert
with the hypothesis of rotational invariance.
That is, the number of events in the $x=+1$ and $x=-1$ channels
only depend on the difference $\theta=\psi-\phi$ between
the (unknown) angle $\psi$ and the orientation of the device $0\le\phi\le2\pi$.
Furthermore, the number of output events should be periodic in
$\theta$ with a period of $\pi$~\cite{known}.

Let the probability $p(x|\theta)$ describe
the process that transforms each input event
into an output event $x=\pm1$.
By symmetry we have $p(x|\theta)=p(x|\theta+\pi)$
and if we assume that each input event generates exactly one
output event we have
\begin{equation}
\sum_{x=\pm1} p(x|\theta)=p(+1|\theta)+p(-1|\theta)=1
.
\label{STOC1}
\end{equation}
We also assume that there is no logical dependence between
two output events $i$ and $j$, that is $p(x_i,x_j|\theta)=p(x_i|\theta)p(x_j|\theta)$
for all $i\not=j$. This implies that the correlation between
the output events is zero.
Then, this process generates Bernoulli trials~\cite{GRIM95,TRIB69,JAYN03}.

Under these conditions, all information about the polarization
is encoded in the measurable quantity
\begin{equation}
f(\theta)=\langle x \rangle=\sum_{x=\pm1} x p(x|\theta)=2p(+1|\theta)-1
.
\label{STOC2}
\end{equation}
From Eq.~(\ref{STOC2}) it is clear that we can completely characterize the process
by $p(\theta)\equiv p(+1|\theta)$.

\setlength{\unitlength}{1cm}

Our task is to design the processor,
that is to determine the function $p(\theta)$,
such that a measurement of $f(\theta)$ gives us as much
as possible knowledge about the unknown angle $\psi$.

\def\psiphi{\theta} 

Let us consider the data as collected by an observer
who decides to record and analyze data sets of $N$ objects each.
We assume that $\psiphi$ is fixed during this measurement.
Each data set looks like $\{x_1,\ldots,x_N\}$ where $x_i=\pm1$ for $i=1,\ldots,N$.
Let us assume that the number of $x_i=+1$ events in a particular data set is $n$.
Recall that the processor generates Bernoulli trials~\cite{TRIB69,GRIM95,JAYN03}.
Therefore, the probability for observing this data set is given by~\cite{TRIB69,GRIM95,JAYN03}
\begin{equation}
P(n|\psiphi,N)=\frac{N!}{n! (N-n)!} p^n(\psiphi) [1-p(\psiphi)]^{N-n}
.
\label{STOC4}
\end{equation}

For convenience of the reader, we first recall some
well known facts of probability theory~\cite{TRIB69,GRIM95,JAYN03}.
From Eq.(\ref{STOC4}) it follows that, as a function of $p(\psiphi)$,
$P(n|\psiphi,N)$ reaches its maximum
$\hat P(n|\psiphi,N)$ at $\hat p(\psiphi)\equiv n/N$.
A simple calculation shows that
\begin{equation}
\frac{1}{N}\ln \frac{P(n|\psiphi,N)}{\hat P(n|\psiphi,N)}=
\hat p(\psiphi)\ln \frac{p(\psiphi)}{\hat p(\psiphi)}
+(1-\hat p(\psiphi))\ln \frac{1-p(\psiphi)}{1-\hat p(\psiphi)}
.
\label{STOC5}
\end{equation}
For small values of $|p(\psiphi)-\hat p(\psiphi)|$,
the Taylor series expansion of the left hand side of Eq.(\ref{STOC5}) yields
\begin{eqnarray}
\frac{P(n|\psiphi,N)}{\hat P(n|\psiphi,N)}
&=&
\exp\left[-N\frac{(p(\psiphi)-\hat p(\psiphi))^2}{2\hat p(\psiphi)(1-\hat p(\psiphi))}
\right]
\nonumber\\
&+&
\ORDER{(p(\psiphi)-\hat p(\psiphi))^4}
,
\label{STOC6}
\end{eqnarray}
showing that as a function of $p(\psiphi)$, $P(n|\psiphi,N)$ vanishes
exponentially fast with $N$, unless $p(\psiphi)=\hat p(\psiphi)=n/N$~\cite{TRIB69,GRIM95}.
Therefore, from the point of view of the observer, the procedure is simple:
As the observer knows (the yet unknown) function $p(\psiphi)$,
after measuring a data set
$\{x_1,\ldots,x_N\}$, the observer finds $\theta$ by solving
$p(\psiphi)=n/N=\left(1+N^{-1}\sum_{i=1}^{N} x_i\right)/2$.
The total number $n$ of $x_i=+1$ events contains all the available information
about the difference $\theta=\psi-\phi$.


A rough estimate for the number of distinguishable
messages $M_D$ that the probabilistic processor can
encode with an error of approximately one percent
can be obtained as follows.
First, we use Eq.~(\ref{STOC4}) to
calculate the variance on $n$ and find
that $\sigma^2(\psiphi)=N\hat p(\psiphi)(1-\hat p(\psiphi))$.
For sufficiently large but fixed $N$,
the probability distribution Eq.~(\ref{STOC4})
tends to the normal (Gaussian) distribution
with mean $n=N p(\psiphi)$ and variance $\sigma(\psiphi)$.
Therefore, the probability to observe $m$ ($1\ll m\ll N$)
instead of $n$ ($1\ll n\ll N$)
$x_i=+1$ events is approximately given by
\begin{eqnarray}
P(m|\psiphi,N)&\approx&
\frac{1}{\sqrt{2\pi \sigma^2(\psiphi)}}
\exp\left[-\frac{(m-n))^2}{2\sigma^2(\psiphi)}
\right]
.
\label{STOC6a}
\end{eqnarray}
From the properties of the normal distribution,
it follows that the probability for the observed number $m$ of
$x_i=+1$ events to lie in the interval
$[n-3\sigma(\psiphi),n+3\sigma(\psiphi)]$ is larger than $0.997$.
Thus, the number of messages $M_D$ that can be
encoded with a probability of error that is less that one
percent is given by
\begin{eqnarray}
M_D&\approx& N/6\sigma(\psiphi)\propto\sqrt{N}
.
\label{MD}
\end{eqnarray}
Although this is a rough estimate,
the result that the number of distinguishable
messages is of the order of $\sqrt{N}$ is to be expected
on general grounds, given the constraint that the processor generates
probabilistic, Bernoulli-like events.

We now apply the criteria 1 and 2 of Section~\ref{sec1}
to optimize the design, that is
we want to determine $p(\theta)$ by which
the probabilistic (Bernoulli) processor will generate
the output events.
In the foregoing analysis,
we assumed that $\psiphi$ was fixed during the
observation of $N$ events.
Clearly, this is not a realistic assumption.
In a real experiment, $\phi$ or $\psi$ fluctuates.
Therefore, the best we can do is search
for the probability $p(\psiphi)$
that is least sensitive to small changes in $\phi-\psi$.
This is criterion 1 of Section~\ref{sec1}.

We determine this probability by considering
the likelihood that the observed sequence of $x_i$'s
was generated by $p(x|\psiphi+\epsilon)$ instead of $p(x|\psiphi)$
where $\epsilon$ is a small positive number.
The larger this likelihood, the larger the probability
that the observer draws the wrong conclusion from the data.
The log-likelihood $L$ that the data was generated
by $p(x|\psiphi+\epsilon)$ instead of by $p(x|\psiphi)$
is given by~\cite{TRIB69,JAYN03}
\begin{eqnarray}
\frac{L}{N}&=&
\frac{1}{N}\ln \frac{P(n|\psiphi+\epsilon,N)}{P(n|\psiphi,N)}
,\nonumber \\
&=&
\frac{n}{N}\ln \frac{p(\psiphi+\epsilon)}{p(\psiphi)}
+(1-\frac{n}{N})\ln \frac{1-p(\psiphi+\epsilon)}{1-p(\psiphi)}
.
\label{STOC5a}
\end{eqnarray}
According to criterion 1 of Section~\ref{sec1}, we have to find
the probability $p(\psiphi)$ that minimizes $|L|$.

We first consider the case that
$P(n|\psiphi+\epsilon,N) > P(n|\psiphi,N)$.
Then, because $P(n|\psiphi,N)$ is not the
maximum, we assign $p(\psiphi+\epsilon)= n/N$
as the most likely guess.
A Taylor series expansion of Eq.~(\ref{STOC5a}) yields
\begin{equation}
\ln \frac{P(n|\psiphi+\epsilon,N)}{P(n|\psiphi,N)}=
\frac{\epsilon^2}{2p(\psiphi)(1-p(\psiphi))}
\left(\frac{\partial p(\psiphi)}{\partial \psiphi}\right)^2
.
\label{STOC7}
\end{equation}
Second, we consider the case that
$P(n|\psiphi+\epsilon,N) \le P(n|\psiphi,N)$.
Now, adopting the same reasoning as used previously, the observer assigns
$p(\psiphi)=n/N$ and the Taylor series expansion of Eq.~(\ref{STOC5a}) yields
\begin{equation}
\ln \frac{P(n|\psiphi+\epsilon,N)}{P(n|\psiphi,N)}=
-\frac{\epsilon^2}{2p(\psiphi)(1-p(\psiphi))}
\left(\frac{\partial p(\psiphi)}{\partial \psiphi}\right)^2
.
\label{STOC8a}
\end{equation}
As $\epsilon$ was arbitrary (but small), minimization of $|L|$ is equivalent
to minimizing the Fisher information~\cite{TREE68,FRIE99,BLAH91}
\begin{equation}
I_F=\frac{1}{p(\psiphi)(1-p(\psiphi))}
\left(\frac{\partial p(\psiphi)}{\partial \psiphi}\right)^2
,
\label{STOC9}
\end{equation}
for this particular problem.
Thus, we conclude that the first criterion tells us
that we should minimize the Fisher information $I_F$.
Substituting $p(\psiphi)=\cos^2 g(\psiphi)$ we obtain
\begin{equation}
I_F=4\left[\frac{\partial g(\psiphi)}{\partial \psiphi}\right]^2
.
\label{STOC10}
\end{equation}

Criterion 2 of Section~\ref{sec1} stipulates that the
reliability of the procedure to extract
$\psi-\phi$ from the observed sequence of $x_i$'s
should not depend on $\psi-\phi$.
We can realize this by choosing $g(\psiphi)=a\psiphi+b$.
Using the side information that $p(+1|\psiphi)=p(+1|\psiphi+\pi)$
we find that $p(+1|\psiphi)=\cos^2(k\psiphi+b)$ and $I_F=4k^2$ for $k\not=0$
($k=0$ is excluded because then $p(\psiphi)$ does not depend
on $\psiphi$ and the design leads to a useless device).
Clearly $I_F$ is minimal if $k=1$
and we may absorb the irrelevant phase factor $b$ in $\phi$.

In summary, using the two design criteria of Section~\ref{sec1},
we find that for optimal operation (from the point of view of the observer),
the processor should use the probabilities
\begin{eqnarray}
p(-1|\psiphi)=\sin^2\psiphi
\quad,\quad
p(+1|\psiphi)=\cos^2\psiphi
,
\label{RATIO}
\label{STOC12}
\end{eqnarray}
to generate the $-1$ and $+1$ events, respectively.
Put differently, for a fixed processor setting $\phi$ and
$N$ incoming events with message $\psi$,
the observer will (in general) get most out of the data
if the processor sends $N \cos^2(\psi-\phi)$
($N \sin^2(\psi-\phi)$)
events to the apparatus that detects
the $+1$ ($-1$) event.
The maximum number $M_D$ of angles $\psi-\phi$
we can distinguish is given by
\begin{eqnarray}
M_D=a\sqrt{N}
,
\label{STOC12a}
\end{eqnarray}
where $a$ depends on
the number of mistakes in determining $\psi-\phi$
that we find acceptable.
The larger $a$, the larger is the probability that the result
for $\psi-\phi$ is erroneous.

Obviously, it is easy to simulate this processor on a computer.
For each of the $i=1,\ldots,N$ input events, we generate a uniform random number $0<r<1$
and send out a $x_i=+1$ ($x_i=-1$) event
if $\cos^2\psiphi\le r$ ($\cos^2\psiphi> r$).
After processing $N$ events,
we compute $\psiphi=\psi-\phi$ from
$\cos^2\psiphi=\left(1+N^{-1}\sum_{i=1}^{N} x_i\right)/2$.

\subsection{Relation to physics}{\label{sec3a}}

Up to this point, there is no relation between
the mathematical model that we have analyzed and a physical system.
However, from the description of the scenario
and the final result Eq.~(\ref{STOC12}),
it is obvious that a processor that operates
according to Eq.~(\ref{STOC12}) is a model for an ideal polarizer.
We now discuss the relation between the optimal
probabilistic processor and the
measurement of the polarization of photons in more detail.

In classical electrodynamics, it is well known
that the intensity of light transmitted
by a polarizer (such as Nicol prism) is given by Malus' law
\begin{eqnarray}
I_o=I\sin^2(\psi-\phi)
\quad,\quad
I_e=I\cos^2(\psi-\phi)
,
\label{STOC13}
\end{eqnarray}
where $I$, $I_o$, and $I_e$ are the intensities
of the incident light, the ordinary and extraordinary ray, respectively,
$\psi$ is the polarization of the incident light
and $\phi$ specifies the orientation of the polarizer~\cite{BORN64}.
From a quantum mechanical point of view,
the total energy $E$ of a light wave of frequency $f$ must
be an integer multiple of $h$ (Planck's constant), that is $E=nhf$, where $n$
is the number of photons in the wave.
The polarizer splits the incoming beam in two beams.
Depending on the type of polarizer, the light in one
of the beams is absorbed~\cite{BORN64} but this is irrelevant
for the discussion that follows.
In any case, the number of photons in each beam is an integer (by
definition of the concept of a photon, there is no such thing as
a half photon).
If the number of photons in the incident beam
is very large, the mean number of photons that goes into each beam
should correspond to the intensity that we find from classical
electrodynamics.
In the regime where the photons are detected one-by-one,
quantum mechanics postulates that the polarizer sends
a photon to the (extra)ordinary direction with
probability ($\sin^2(\psi-\phi)$) $\cos^2(\psi-\phi)$~\cite{BAYM74}.

The probabilistic processor that we have described
transforms a beam of photons into yes/no events that we can count.
If we require the answers of the transformation process
to be probabilistic (Bernoulli trials), rotational invariant (a basic property
of (quantum) electrodynamics),
and to satisfy criteria 1 and 2 of Section~\ref{sec1},
then the device that performs the transformation will
produce data that agrees with Malus' law.
We did not invoke any law of physics to obtain this result:
Malus' law was recovered as the result of efficient data processing.
This raises the interesting question whether other quantum phenomena
also appear as the result of efficient data processing.

The hypothesis that efficient processing of
statistical information may be the reason why we observe
quantum mechanical phenomena is very explicit in the work of
Frieden~\cite{FRIE99}, Wootters~\cite{WOOT81}, and Summhammer~\cite{SUMM01}.
Frieden has shown that one can recover all the fundamental equations of
physics by finding the extrema of the Fisher information plus
the ``bound'' information~\cite{FRIE99}.
According to Frieden, the act of measurement elicits a physical law
and quantum mechanics appears as the result of what Frienden calls "a smart measurement",
a measurement that tries to make the best estimate~\cite{FRIE99}.
Although this approach is similar to ours,
our line of reasoning is different.
We do not invoke concepts from estimation theory, such as the
estimators and the Cram\'er-Rao inequality (see Appendix A), nor do we require
the concept of random noise. Furthermore, in Frieden's
formulation, the parameters to be
estimated (such as the position) are of the same kind
as the measured quantities.
This is not the case for the photon polarization that we treat here.
In our approach, the measuring apparatus (such as the calcite crystal acting
as a polarizer) transforms the input (the photon polarization)
into a signal ($x=\pm1$) that can be detected by human beings.
The requirement that the simple probabilistic processor, that
transforms the data, operates with optimal efficiency
yields Malus' law.

The fundamental difference between Frienden's approach and ours
becomes evident by noting that there is no reason
why we should limit our search for efficient transformation
devices to the most simple, Bernoulli-type probabilistic machines.
As we explain later, these machines can simulate the classical
and quantum properties of a photon polarizer but are
incapable of simulating interference phenomena.
One possible route to solve this problem
might be to generalize the probabilistic machine
such that it no longer generates Bernoulli events,
that is allow for correlations between output events.
We don't follow this route.
Instead we consider the most extreme solution, namely
a deterministic processor that
performs the same task as the probabilistic machine
under the conditions specified in Section \ref{sec1}.
This forces us to consider deterministic algorithms
with primitive learning capabilities
(to allow for correlations between output events).
Elsewhere, we have shown that these
deterministic processors (and probabilistic versions thereof)
can be used to reproduce quantum interference
phenomena~\cite{KRAED05,HRAED05a,MICH05,HRAED05b}.
We come back to this topic in Section~\ref{sec5}.

\section{Deterministic processor}{\label{sec3}}

From an engineering point of view, the probabilistic processor
of Section~\ref{sec2} is extremely simple
and has a relatively poor performance.
Using $N$ bits, the probabilistic processor can
encode $M_D\propto\sqrt{N}$ distinguishable messages only.
For example, as shown in Section~\ref{sec2}, if we demand
the level of certainty of 99.7\%, then $M_D\approx\sqrt{N}/6$.

It is not unreasonable to expect that a deterministic machine
can do better in this respect.
Therefore, the obvious question is to ask if there exists
a deterministic processor that generates events according to Malus' law.
Apart from being deterministic, this processor
should satisfy the two criteria that we
specified in Section~\ref{sec1}.

Adopting the terminology introduced in our
earlier work~\cite{KRAED05,HRAED05a,MICH05,HRAED05b},
we refer to the deterministic processor that
we describe in this section as a deterministic learning machine (DLM).
For this machine, $M_D=N+1$ with nearly 100\% certainty.

In this paper, we analyze a \DLM\ that has one input channel,
two output channels and one internal vector with two real entries.
A \DLM\ responds to an input event by
choosing from all possible alternatives, the internal state
that minimizes a cost function (to be defined later) that depends on
the input and the internal state itself.
Then the \DLM\ sends a message through one of its output channels.
The message contains information about the decision the \DLM\ took
while it updated its internal state and, depending on the application,
also contains other data that the \DLM\ may have.
By updating its internal state, the \DLM\ ``learns" about the input it receives
and by sending messages through one of its two output channels, it tells
its environment about what it has learned.
A \DLM\ is a machine that performs real-time recurrent learning~\cite{HAYK99}.

This section consists of three parts.
First, we specify the algorithm that is used by a \DLM\
and we show that in the stationary regime,
the number of $-1$ ($+1$) events in a sequence of $N$
events is given by Malus' law, see Eq.~(\ref{STOC13}).
Then, we present a detailed mathematical analysis
of the dynamic properties of a \DLM.
The reader who is not interested in the intricacies
of this classical dynamical system
can skip Section~\ref{sec4b}.
We end this section by comparing the
performance of the probabilistic
and deterministic processor.

\subsection{Deterministic Learning Machines}{\label{sec4a}}

The schematic diagram of the \DLM\ is the same as
that of the probabilistic processor of Fig.~\ref{machine1},
except that there is no probabilistic process $p(x|\psi-\phi)$.
The \DLM\ receives as input, a sequence of angles $\psi_{n+1}$ for $n=0,\ldots,N$
and also knows about the orientation of the device through the angle $\phi$.
Using rotational invariance, we represent these input messages
by unit vectors ${\bf y}_{n+1}=(y_{1,n+1},y_{2,n+1})$  where
\begin{equation}
y_{1,n+1}=\cos \psiphi_{n+1}\quad y_{2,n+1}=\sin \psiphi_{n+1}
,
\label{NDIMangle}
\end{equation}
and $\psiphi_n=\psi_n-\phi$.
The fact that Eq.~(\ref{NDIMangle})
depends on the relative difference of the angles
guarantees that the deterministic process
is rotational invariant.
Instead of the random number generator that is
part of the probabilistic processor,
the \DLM\ has an internal degree of freedom that we represent
by the unit vector ${\bf x}_{n+1}=(x_{1,n+1},x_{2,n+1})$.
As the \DLM\ receives input data, it updates its internal state.
For all $n>0$, the update rules are defined by
\begin{eqnarray}
x_{1,n+1}&=&
\alpha x_{1,n} + \beta(1-\Theta_{n+1})
,\nonumber \\
x_{2,n+1}&=&
\alpha x_{2,n} + \beta\Theta_{n+1},
\label{CIRC1}
\end{eqnarray}
where $\Theta_{n+1}=0$ ($1$) corresponds to an $-1$ ($+1$) output event,
and $0<\alpha<1$ is a parameter that controls the learning process of the \DLM.
The requirement that the internal vector
${\bf x}_{n+1}=(x_{1,n+1},x_{2,n+1})$ stays on the unit circle yields
\begin{eqnarray}
\beta=
&\pm&\sqrt{1+\alpha^2[x_{1,n}^2(1-\Theta_{n+1}) +x_{2,n}^2\Theta_{n+1}-1]}
\nonumber \\
&-&\alpha[x_{1,n}(1-\Theta_{n+1}) +x_{2,n}\Theta_{n+1}]
.
\label{CIRC2}
\end{eqnarray}
Substitution of Eq.~\Eq{CIRC2} in Eq.~\Eq{CIRC1} gives us
four different rules:
\begin{eqnarray}
x_{1,n+1}=&+\sqrt{1+\alpha^2(x_{1,n}^2-1)}
,\quad
x_{2,n+1}=\alpha x_{2,n}
,\nonumber \\
x_{1,n+1}=&-\sqrt{1+\alpha^2(x_{1,n}^2-1)}
,\quad
x_{2,n+1}=\alpha x_{2,n}
,\nonumber \\
x_{1,n+1}=&\alpha x_{1,n}
,\quad
x_{2,n+1}=+\sqrt{1+\alpha^2(x_{2,n}^2-1)}
,\nonumber \\
x_{1,n+1}=&\alpha x_{1,n}
,\quad
x_{2,n+1}=-\sqrt{1+\alpha^2(x_{2,n}^2-1)}
,
\label{CIRC3}
\end{eqnarray}
where the first (last) two rules correspond to the
choice $\Theta_{n+1}=0$ ($\Theta_{n+1}=1$)
and the $\pm$-sign takes care of the fact that
for each choice of $\Theta_{n+1}$, the \DLM\ has to decide between two quadrants.
For later, it is important to note that
$|x_{1,n+1}|>|x_{1,n}|$ and $|x_{2,n+1}|<|x_{2,n}|$ if $\Theta_{n+1}=0$.
In other words, the angle of the internal vector relative to
the $x$-axis decreases if we apply the $\Theta_{n+1}=0$ rules.
The \DLM\ selects one of the four rules in Eq.~(\ref{CIRC3}) by
minimizing the cost function defined by
\begin{eqnarray}
C&=&-{\bf x}_{n+1}\cdot{\bf y}_{n+1}=({\bf x}_{n+1}-{\bf y}_{n+1})^2/2-1
\nonumber \\
&=&-(x_{1,n+1}y_{1,n+1}+x_{2,n+1}y_{2,n+1})
.
\label{CIRC4}
\end{eqnarray}
Obviously, the cost $C$ is small if
the vectors ${\bf x}_{n+1}$ and ${\bf y}_{n+1}$ are close to each other.
Summarizing: a \DLM\ minimizes the distance between the input vector
and its internal vector by means of a simple, deterministic decision process.

\setlength{\unitlength}{1cm}
\begin{figure}[t]
\begin{center}
\includegraphics[width=8cm]{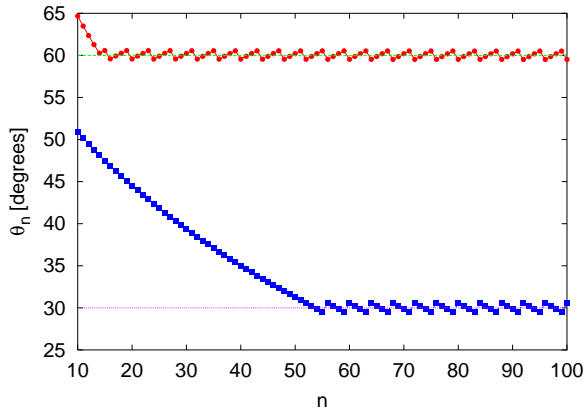}
\caption{%
(color online) Time evolution of the angle $\theta_n=\arctan(x_{2,n}/x_{1,n})$
representing the internal vector ${\bf x}_{n}$
of the \DLM\ defined by Eqs.~\Eq{CIRC3} and \Eq{CIRC4}.
Bullets (red): Input events carry vectors ${\bf y}_{n+1}=(\cos 60^\circ,\sin 60^\circ)$.
The initial value $\theta_0\approx81^\circ$.
For $n>20$ the ratio of the number of increments ($\Theta_{n+1}=1$)
to decrements ($\Theta_{n+1}=0$) is exactly 3/1, which is $(\sin 60^\circ/\cos 60^\circ)^2$.
Squares (blue): Input events carry vectors ${\bf y}_{n+1}=(\cos 30^\circ,\sin 30^\circ)$.
The initial value $\theta_0\approx327^\circ$.
For $n>60$ the ratio of the number of increments ($\Theta_{n+1}=1$)
to decrements ($\Theta_{n+1}=0$) is exactly 1/3,
which is $(\sin 30^\circ/\cos 30^\circ)^2$.
The direction of the initial vectors ${\bf x}_{0}$ is chosen at random.
In this simulation $\alpha=0.99$.
Data for $n<10$ has been omitted to show the oscillating behavior more clearly.
Lines are guides to the eyes.
}
\label{c60}
\end{center}
\end{figure}
\begin{figure}[t]
\begin{center}
\includegraphics[width=8cm]{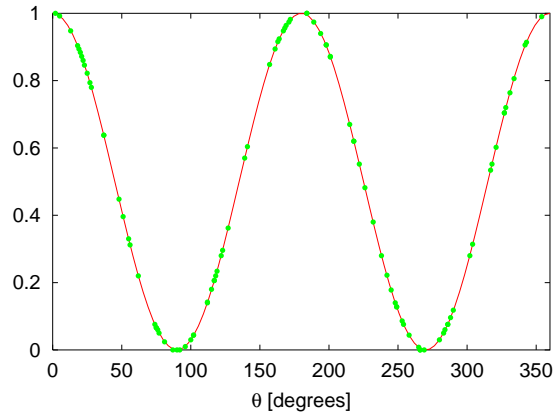}
\caption{%
The number of ($\Theta_{n+1}=1$) events
divided by the total number of events
as a function of the value of the input variable $\psiphi$.
Bullets: Each data point is obtained from a \DLM\
simulation of 1000 events with a fixed, randomly
chosen value of $0\le\phi<360^\circ$, using the last 500 events
to count the number of ($\Theta_{n+1}=1$) events.
Solid line: $\cos^2\psiphi$.
}
\label{circle2}
\end{center}
\end{figure}

In general, the behavior of the \DLM\ defined by rules
Eqs.~\Eq{CIRC3} and \Eq{CIRC4} is difficult to analyze without the use of a computer.
However, for a fixed input vector ${\bf y}_{n+1}={\bf y}$,
it is clear what the \DLM\ will try to do:
It will minimize the cost Eq.~\Eq{CIRC4} by rotating
its internal vector ${\bf x}_{n+1}$ to bring it as close as possible to ${\bf y}$.
However, ${\bf x}_{n+1}$ will not converge to a limiting value but instead
it will keep oscillating about the input value ${\bf y}$.
An example of a simulation is given in Fig.~\ref{c60}.
In general, for a fixed input vector ${\bf y}_{n+1}={\bf y}$
the \DLM\ will reach a state
in which its internal vector oscillates about ${\bf y}$.
This is the stationary state of the machine.
Obviously, the whole process is deterministic.
The details of the approach to the stationary state depend on
the initial value of the internal vector ${\bf x}_0$,
but the properties of the stationary state do not.

\subsubsection{Stationary state}{\label{station}}

The stationary-state analysis is a very
useful tool to understand the behavior of the \DLMS.
Let us assume that $0\ll \alpha<1$ and that we have reached
the stationary regime in which the internal vector performs
small oscillations about $(\cos\psiphi,\sin\psiphi)$ (as in Fig.~\ref{c60}).
For simplicity, but without loss of generality, we
limit the discussion that follows to $0\le \psiphi\le\pi/2$.
For $\Theta_{n+1}=0$
we substitute $x_{2,n}=\sin\varphi_{n}$
and $\psiphi_{n+1}=\varphi_{n}+\delta_0$
in Eq.~(\ref{CIRC3}) and obtain
\begin{eqnarray}
\sin^2\varphi_n+2\delta_0\sin\varphi_n\cos\varphi_n&=&\alpha^2\sin^2\varphi_n
.
\label{CIRC5b0}
\end{eqnarray}
Similarly, for $\Theta_{n+1}=1$
we substitute $x_{2,n}=\sin\varphi_{n}$
and $\psiphi_{n+1}=\varphi_{n}+\delta_1$
in Eq.~(\ref{CIRC3}) and obtain
\begin{eqnarray}
\sin^2\varphi_n+2\delta_1\sin\varphi_n\cos\varphi_n
&=&-\alpha^2\cos^2\varphi_n+1
.
\label{CIRC5b1}
\end{eqnarray}
In deriving Eqs.~(\ref{CIRC5b0}) and ~(\ref{CIRC5b1}),
we neglected terms of order $\delta^2_0$ and $\delta^2_1$,
respectively.
Rearranging Eqs.~(\ref{CIRC5b0}) and ~(\ref{CIRC5b1}),
and using $\varphi_n\approx\psiphi$ gives
\begin{eqnarray}
\delta_0&=&-\frac{1-\alpha^2}{2}\frac{\sin\psiphi}{\cos\psiphi}
\quad\hbox{if\ \ } \Theta_{n+1}=0
,\nonumber \\
\delta_1&=&\phantom{-}\frac{1-\alpha^2}{2}\frac{\cos\psiphi}{\sin\psiphi}
\quad\hbox{if\ \ } \Theta_{n+1}=1
.
\label{CIRC5a}
\end{eqnarray}
In the stationary regime, the sum of all increments of $\varphi_n$
should be compensated by the sum of all decrements of $\varphi_n$.
Therefore, we must have $N_0\delta_0+N_1\delta_1\approx 0$
where $N_0$ ($N_1$) is the number of $\Theta_{n+1}=0$ ($\Theta_{n+1}=1$) events.
From Eq.~\Eq{CIRC5a} it follows immediately that
\begin{equation}
\tan^2\psiphi\approx\frac{N_1}{N_0}
,
\label{RATIO2}
\end{equation}
and hence
\begin{eqnarray}
\frac{N_1}{N_0+N_1}\approx \sin^2\psiphi
\quad,\quad
\frac{N_0}{N_0+N_1}\approx \cos^2\psiphi
.
\label{CIRC5c}
\end{eqnarray}
Fig.~\ref{circle2} shows that
the simulation results generated by the \DLM\
are in excellent agreement with
the expressions obtained from this simple analysis.
In fact, we will see later that
in the stationary state, a \DLM\ can encode exactly
all angles for which $\sin^2\theta=n/N$ where $n=0,\ldots,N$.
From the definition of the \DLM\ algorithm and
Eq.~\Eq{CIRC5c}, it is clear that
the requirements of rotational invariance and insensitivity with respect
to small changes in $\theta=\phi-\psi$ (criterion 1 of Section~\ref{sec1})
are automatically satisfied.
We emphasize that Eq.~\Eq{CIRC5c} is not put into the \DLM\ algorithm
but results from the learning process itself.

Comparing Eq.~\Eq{STOC13} and Eq.~\Eq{CIRC5c},
we conclude that once the \DLM\ has reached a stationary state,
the number of $+1$ and $-1$ output events
in a sequence of $N=N_0+N_1$ events agrees with Malus' law.
Of course, the order in which the
\DLM\ generates the $+1$ and $-1$ is strictly deterministic.
Anticipating that we will show that a \DLM\ is
a very efficient machine, what is most striking is
that the number of $-1$ and $+1$ events it generates is
proportional to $\sin^2(\psi-\phi)$
and $\cos^2(\psi-\phi)$, respectively, just
as in the case of the simple probabilistic processor
and in the classical electrodynamical and quantum mechanical
description of the polarizer.

\subsection{Analysis of the dynamic properties}{\label{sec4b}}

For a more detailed mathematical analysis of the dynamics of a \DLM,
it is convenient to write the update rules Eq.~\Eq{CIRC3} as linear difference equations.
Actually, we need only
\begin{eqnarray}
x_{2,n+1}^2&=&\alpha^2 x_{2,n}^2 + (1-\alpha^2)\Theta_{n+1}.
\label{CIRC5}
\label{RECURSION}
\end{eqnarray}
For simplicity, we restrict the discussion that follows to
the case $0\le \psiphi\le\pi/4$.
Other cases can be treated in the same manner.

\begin{figure}[t]
\begin{center}
\includegraphics[height=7cm]{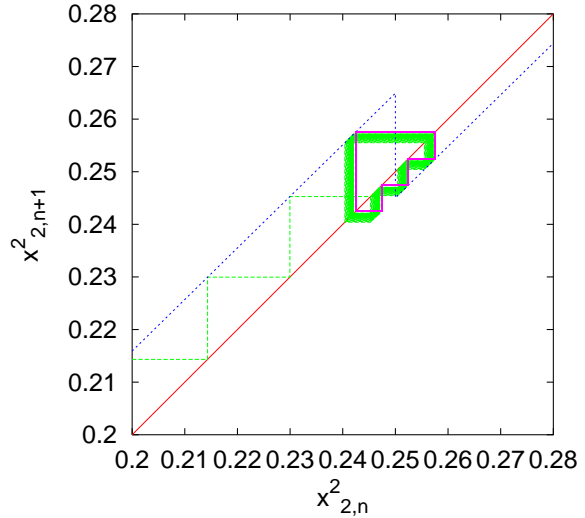}
\caption{(color online)
Circle map of the time evolution of $x^2_{2,n}$
for the case of a fixed input angle of $30^\circ$.
The dashed (green) line shows the evolution
of the mapping $x^2_{2,n+1}=F(x^2_{2,n})$ for $n<100$.
For clarity, we omitted the first 12 iterations
because this allows us to show in detail how the mapping converges
to a unique polygon.
The function $F(x^2)$ is defined by the rules and
cost function Eq.~(\ref{CIRC3}) and
Eq.~(\ref{CIRC4}), respectively.
The dotted (blue) line separates the case $\Theta_{n+1}=0$ from
the case $\Theta_{n+1}=1$ and is given by
$y=\alpha^2 x + 1-\alpha^2$
for $\Theta_{n+1}=1$ ($x<1/4$)
and $y=\alpha^2 x$
for $\Theta_{n+1}=0$ ($x>1/4$).
The straight solid (red) line is given by $y=x$.
The solid (red) line forming the polygon with eight vertices
shows the results for $9900\le n<10000$: In this case the
system has reached the stationary state with a period of four.
In this simulation, $\alpha=0.99$.
}
\label{figcirclemap}
\end{center}
\end{figure}

Substituting $x_{2,n}=\sin\varphi_{n}$ in Eq.~(\ref{RECURSION}), we obtain
\begin{eqnarray}
\sin^2\varphi_{n+1}&=&\alpha^2 \sin^2\varphi_{n} + (1-\alpha^2)\Theta_{n+1},
\label{CIRCLEMAP}
\end{eqnarray}
showing that Eq.~(\ref{RECURSION})
has the structure of a so-called circle map~\cite{JENS83}.
Thus, the study of the behavior of the circle map
Eq.~(\ref{CIRCLEMAP}) will give us insight into
the dynamic properties of the \DLM.
Fig.~\ref{figcirclemap} shows an example of circle-map
analysis for the case of a fixed input angle of $30^\circ$.

\subsubsection{Illustrative example}{\label{pattern}}

Let us assume that we have reached a stationary state and
that the \DLM\ repeats a sequence
$\{00\ldots00100\ldots00100\ldots\}$
in which there are $K$ successive events of the type $\Theta_{n+1}=0$
(decreasing $x_{2,j}$) and one $\Theta_{n+1}=1$ event (increasing $x_{2,j}$).
Let us denote by $\hat x$, the value of $x_{2,n+1}$
before the first of the $K$ events of type 0.
From Eq.~\Eq{CIRC5} we obtain
\begin{eqnarray}
x_{2,K}^2&=&\alpha^{2K} \hat x^2 
,\nonumber \\
x_{2,K+1}^2&=&\alpha^{2} x_{2,K}^2+ 1-\alpha^{2}
,\nonumber \\
&=&\alpha^{2K+2} \hat x^2 + 1-\alpha^{2} 
\label{CIRC6a}
\end{eqnarray}
As the \DLM\ repeats the same sequence over and over again,
we have $x_{2,K+1}^2=\hat x^2$.
In other words, if we observe the repeated sequence $\{00\ldots01\}$ of length $K+1$,
we must have
\begin{eqnarray}
\hat x^2&=&\frac{1-\alpha^{2}}{1-\alpha^{2K+2}}.
\label{CIRC6b}
\end{eqnarray}
Furthermore, as $x_{2,j}^2=\alpha^{2j} \hat x^2$ for $j=0,\ldots,K$,
the mean value of the $x^2_{2,j}$'s during the sequence is given by
\begin{eqnarray}
\EXPECT{x^2}&\equiv&\frac{1}{K+1}\sum_{j=0}^K x_{2,j}^2=\frac{1}{K+1}\approx\sin^2\theta
,
\label{CIRC6c}
\end{eqnarray}
in agreement with Eq.~\Eq{CIRC5c}.
From Eq.~\Eq{CIRC6c}, we conclude that the \DLM\ can encode
the values $\psiphi=\arctan(1/\sqrt{K})$ with periodic sequences of the
form $\{00\ldots01\}$.

From this analysis we conclude that if we would
limit the design of the device such that it can only
generate sequences of the form $\{00\ldots01\}$,
then, after observing two one's and counting the zeros between
these two one's, we can determine the angle with an error of less than
5 degrees. This is the worst case and occurs when the sequence
is $\{010101010\ldots\}$ ($45\deg$) and $\{0010010010\ldots\}$ ($35\deg$).
Clearly, even with this limitation ($K$ zero's followed by one $1$)
on the design, this is already a very efficient method to encode the angle.

We now extend this analysis to a general periodic sequence.

\subsubsection{Minimum angle}{\label{resol}}

\begin{figure}[t]
\begin{center}
\includegraphics[width=7.5cm]{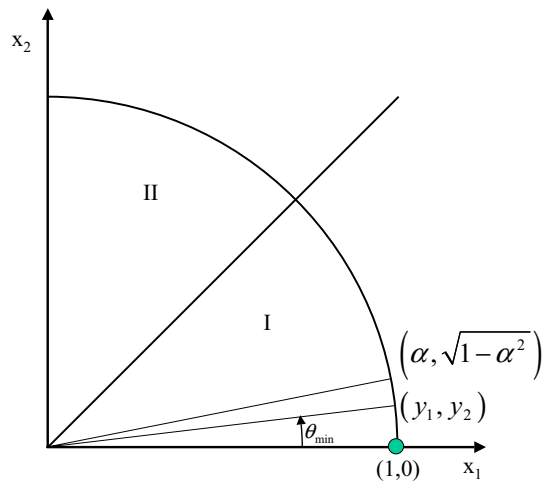}
\caption{(color online) Illustration of a situation
in which the machine remains in the state ${\bf x}=(1,0)$.
The input vector is ${\bf y}=(y_1,y_2)$.
The internal state is ${\bf x}_{n}=(1,0)$, and the
new internal state is either
${\bf x}_{n+1}=(\alpha,\sqrt{1-\alpha^2})$
or
${\bf x}_{n+1}=(1,0)$.
In general, the smallest angle $\theta_{min}$
for which the machine remains in the state ${\bf x}=(1,0)$
depends on the value of the parameter $\alpha$, see Eq.~(\ref{ALPH5}).
}
\label{machine3}
\end{center}
\end{figure}

First we show how the control parameter $\alpha$ limits the accuracy
with which we can represent the stationary state.
Let us assume that the fixed input vector is
given by ${\bf y}=(y_1,y_2)$ and that for some index $n$,
the machine is in the state ${\bf x}_{n}=(1,0)$,
as illustrated in Fig.~\ref{machine3}
(the cases $(0,1)$, $(-1,0)$, and $(0,-1)$ can be treated in the same
manner and lead to the same conclusion).

If the machine applies the update rule $\Theta_{n+1}=0$,
the new state and the cost are given by
\begin{eqnarray}
x_{1,n+1}&=&\sqrt{1+\alpha^2(x_{1,n}^2-1)}=1
,\nonumber \\
x_{2,n+1}&=&\alpha x_{2,n}=0
,\nonumber \\
C&=&-y_1.
\label{ALPH2}
\end{eqnarray}
The cost $C$ has to be compared to the cost
of applying the update rule $\Theta_{n+1}=1$,
in which case we have
\begin{eqnarray}
x_{1,n+1}&=&\alpha x_{1,n}=\alpha
,\nonumber \\
x_{2,n+1}&=&\sqrt{1+\alpha^2(x_{2,n}^2-1)}=\sqrt{1-\alpha^2}
,\nonumber \\
C&=&-(\alpha y_1+y_2\sqrt{1-\alpha^2}).
\label{ALPH1}
\end{eqnarray}

Note that the point $(1,0)$ is somewhat special
in the sense that the machine remains at $(1,0)$
if it applies the update rule $\Theta_{n+1}=0$.
The machine stays at $(1,0)$ (forever) unless
the cost of applying the update rule $\Theta_{n+1}=1$,
is less than the cost of applying the update rule $\Theta_{n+1}=0$.
From Eqs.~(\ref{ALPH2}) and (\ref{ALPH1}), the necessary
condition for the machine not to get stuck at $(1,0)$ is
\begin{eqnarray}
\alpha y_1+y_2\sqrt{1-\alpha^2} > y_1.
\label{ALPH3}
\end{eqnarray}
Rearranging Eq.~(\ref{ALPH3}) yields
\begin{eqnarray}
\tan^2\theta=\frac{y_2^2}{y_1^2} > \frac{1-\alpha}{1+\alpha}.
\label{ALPH4}
\end{eqnarray}
Thus, Eq.~(\ref{ALPH1}) shows that we
cannot represent angles $\theta$ that are smaller than
\begin{eqnarray}
\theta_{min}&=&\arctan\sqrt{(1-\alpha)/(1+\alpha)}.
\label{ALPH5}
\end{eqnarray}
For $\alpha=0.99$ ($0.999$), typical values used in simulations,
$\theta_{min}=4.05^\circ$ ($1.28^\circ$).
Note that $\theta_{min}$ does not determine the accuracy
in the interval $[\theta_{min},\pi/4]$.

\subsubsection{Periodic sequences: General case}{\label{fixed}}

We now consider situations in which
the sequence of events consists of a repetition of
the sequence $\{\Theta_{n+1},\Theta_{n+2},\ldots,\Theta_{n+N}; \Theta_{n}=\Theta_{n+N}\}$
of length $N$.
First, we determine the solution
$\hat x_{2,n}^2$ of $x_{2,n}^2=x_{2,n+N}^2$
(implying $x_{1,n}^2=x_{1,n+N}^2$).
The formal solution of Eq.~(\ref{CIRC5}) is given by
\begin{eqnarray}
x_{2,n+k}^2&=&\alpha^{2k} x_{2,n}^2
 + (1-\alpha^2)\sum_{j=1}^k \alpha^{2(k-j)}\Theta_{n+j},
\label{FIX1}
\end{eqnarray}
and the requirement $x_{2,n}^2=x_{2,n+N}^2$ yields
\begin{eqnarray}
{\hat x}_{2,n+N}^2&=&\frac{1-\alpha^{2}}{1-\alpha^{2N}}
\sum_{j=1}^N \alpha^{2(N-j)}\Theta_{n+j}.
\label{FIX1a}
\end{eqnarray}
We conclude that if the machine starts from $\hat x_{2,n}^2$
and generates the events
$\{\Theta_{n+1},\Theta_{n+2},\ldots,\Theta_{n+N}\}$,
it returns to the starting point $\hat x_{2,n}^2$.
For each pattern $\{\Theta_{n+1},\Theta_{n+2},\ldots,\Theta_{n+N}\}$,
there exists such a point $\hat x_{2,n}^2$.
In other words, if the machine is in the state $\hat x_{2,n}^2$,
repeating the sequence $\{\Theta_{n+1},\Theta_{n+2},\ldots,\Theta_{n+N}\}$
generates a periodic motion of $x_{2,n+k}^2$ for $k>0$ with period $N$.

Second, we consider the situation in which
the machine starts from $\hat x_{2,n}^2+\epsilon$
and we keep feeding the machine with the periodic sequence
$\{\Theta_{n+1},\Theta_{n+2},\ldots,\Theta_{n+N}\}$.
Using the general expression Eq.~(\ref{FIX1}), we find
%
%
\begin{eqnarray}
x_{2,n+pN}^2&=&\alpha^{2pN} \hat x_{2,n}^2 +\alpha^{2pN}\epsilon
\nonumber\\
 &&+ (1-\alpha^2)\sum_{j=1}^{pN} \alpha^{2(pN-j)}\Theta_{n+j},
\nonumber\\
&=&\alpha^{2N} \hat x_{2,n+(p-1)N}^2 +\alpha^{2pN}\epsilon
\nonumber\\
&& + (1-\alpha^2)\sum_{j=1}^{N} \alpha^{2(N-j)}\Theta_{n+j},
\nonumber\\
&=&\alpha^{2N} \hat x_{2,n}^2 +\alpha^{2pN}\epsilon
\nonumber\\
&& + (1-\alpha^2)\sum_{j=1}^{N} \alpha^{2(N-j)}\Theta_{n+j},
\label{FIX1b}
\end{eqnarray}
where $p$ denotes the number of times the machine
processes the periodic sequence
$\{\Theta_{n+1},\Theta_{n+2},\ldots,\Theta_{n+N}\}$.
As $\alpha<1$, $\lim_{p\rightarrow\infty} x_{2,n+pN}^2=\hat x_{2,n}^2$,
independent of the choice of $\epsilon$.
Therefore, for any periodic sequence
$\{\Theta_{n+1},\Theta_{n+2},\ldots,\Theta_{n+N}; \Theta_{n}=\Theta_{n+N}\}$
of length $N$, the corresponding sequence
$\{x_{2,j+1}^2,x_{2,j+2}^2,\ldots,x_{2,j+N}^2\}$
converges exponentially fast to the periodic sequence
$\{\hat x_{2,j+1}^2,\hat x_{2,j+2}^2,\ldots,\hat x_{2,j+N}^2\}$.
From Eq.~(\ref{CIRC5c}) it then follows that
\begin{eqnarray}
\frac{1}{N}\sum_{i=0}^{N-1}\hat x_{2,n+1+i}^2&=&
\frac{\alpha^2 }{N}\sum_{i=0}^{N-1}\hat x_{2,n+i}^2
\nonumber\\
&&+
\frac{1-\alpha^2 }{N}\sum_{i=0}^{N-1}\Theta_{n+1+i}.
\label{FIX2c}
\end{eqnarray}
and using $\hat x_{2,n+N}=\hat x_{2,n}$ we find
\begin{eqnarray}
\frac{1}{N}\sum_{j=1}^N
\hat x_{2,n+j}^2
&=&
\frac{1}{N}\sum_{j=1}^N \Theta_{n+j}
\equiv {\bar \Theta}
.
\label{FIX2a}
\end{eqnarray}
Note that ${\bar \Theta}$ is a rational number
and that according to Eq.~(\ref{CIRC5c}),
we have ${\bar \Theta}\approx\sin^2\theta$.

\subsubsection{Lowerbound on the control parameter $\alpha$}{\label{lower}}

Previously, we have tactically assumed that we can always
find the periodic sequence of $\Theta_n$'s that represents the input angle $\theta$.
We now show that for a fixed input angle $\theta$,
the control parameter $\alpha$ has to be large enough (but smaller than one)
in order that the \DLM\ repeats the same
sequence $\{\Theta_{n+1},\Theta_{n+2},\ldots,\Theta_{n+N}\}$.
As before, we confine the discussion to input angles that satisfy $0\le\tan\theta=y_2/y_1\le1$.
Then, the number of 0 events is larger than the number of 1 events.
Without loss of generality, we may put $\Theta_{n+1}=0$.
This means that the internal state $(x_{1,n},x_{2,n})$ of the \DLM\ satisfies $x_{2,n}>y_2$.
If the sequence is to be periodic with period $N$, we must have $x_{2,n+N}=x_{2,n}$.

\begin{figure}[t]
\begin{center}
\includegraphics[width=8.5cm]{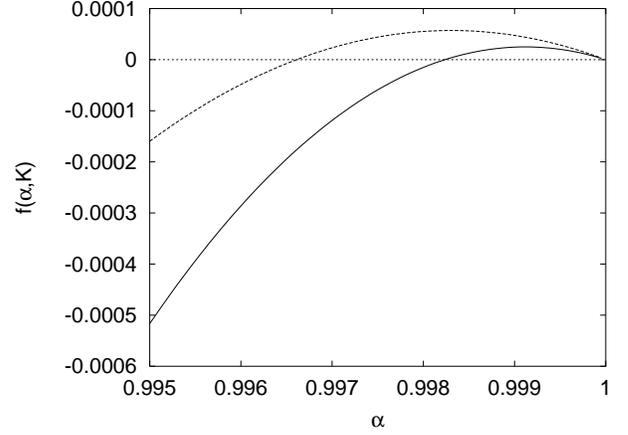}
\caption{Plot of the function $f(\alpha,K)$ (see Eq.~(\ref{CRIT4}))
as a function of $\alpha$ for $K=57$ (solid line) and $K=80$ (dashed line).
In the stationary regime and for $f(\alpha,K)>0$,
the \DLM\ repeats the sequence $\{00\ldots001\}$ with $K$ zero's,
that is, it generates and exact representation of $K$.
}
\label{alpha}
\end{center}
\end{figure}

%
%
\begin{table*}[t]
\caption{
Sequences $\{\Theta_1,\ldots,\Theta_q\}$
marked with a $^\ast$ yield the smallest variance $\Delta^2$.
}
\begin{center}
\begin{ruledtabular}
\begin{tabular}{clcc}
$\bar\Theta=p/q$ & $\{\Theta_1,\ldots,\Theta_q\}$ & $\hat x_{2,0}^2$ & $\Delta^2$\\
\hline
\noalign{\vskip 4pt}
$1/2$ & 10$^\ast$ & $\frac{\alpha^{2}}{{1 + \alpha^{2}}}$ & $\frac{{\left( 1 - \alpha^2 \right) }^2}{4\,{\left( 1 + \alpha^2 \right) }^2}$ \\
\noalign{\vskip 2pt}
\hline
\noalign{\vskip 2pt}
$1/3$ & 100$^\ast$ & $\frac{\alpha^{4}}{{{1 + \alpha^{2} + \alpha^{4}}}}$ & $\frac{2\,{\left( 1 - \alpha^2 \right) }^2}{9\,\left( 1 + \alpha^2 + \alpha^4 \right) }$\\
\noalign{\vskip 2pt}
\hline
\noalign{\vskip 2pt}
$1/4$ & 1000$^\ast$ &
$\frac{\alpha^6}{(1+ \alpha^2)(1 + \alpha^{4})}$ & $\frac{{\left( 1 - \alpha^2 \right) }^2\,\left( 3 + 4\,\alpha^2 + 3\,\alpha^4 \right) }{16\,{\left( 1 + \alpha^2 \right) }^2\,\left( 1 + \alpha^4 \right) }$ \\
\noalign{\vskip 2pt}
\hline
\noalign{\vskip 2pt}
$2/5$ &11000 & $\alpha^6\,{{\frac{1 - \alpha^4}{1 - \alpha^{10}}}}$ & $\frac{2\,{\left( 1 - \alpha^2 \right) }^2\,\left( 3 + 4\,\alpha^2 + 3\,\alpha^4 \right) }{25\,\left( 1 + \alpha^2 + \alpha^4 + \alpha^6 + \alpha^8 \right) }$ \\
$2/5$ &10100$^\ast$ & $\alpha^8\,{{\frac{(1+\alpha^4)(1-\alpha^2)}{1 - \alpha^{10}}}}$ & $\frac{2\,{\left( 1 - \alpha^2 \right) }^2\,\left( 3 - \alpha^2 + 3\,\alpha^4 \right) }{25\,\left( 1 + \alpha^2 + \alpha^4 + \alpha^6 + \alpha^8 \right) }$ \\
\noalign{\vskip 2pt}
\hline
\noalign{\vskip 2pt}
$2/8$ &11000000 & $\alpha^{12}\,\frac{1- \alpha^4}{1 - \alpha^{16}}$ & $\frac{{\left( 1 - \alpha^2 \right) }^2\,\left( 3 + 2\,\alpha^2 + 4\,\alpha^4 + 2\,\alpha^6 + 3\,\alpha^8 \right) }{16\,\left( 1 + \alpha^4 + \alpha^8 + \alpha^{12} \right) }$ \\
$2/8$ &10100000 & $\alpha^{10}\,\frac{1}{{{1 + \alpha^{2} + \alpha^{8} + \alpha^{10}}}}$ & $\frac{{\left( 1 - \alpha^2 \right) }^2\,\left( 3 + 4\,\alpha^2 + 4\,\alpha^4 + 4\,\alpha^6 + 3\,\alpha^8 \right) }{16\,{\left( 1 + \alpha^2 \right) }^2\,\left( 1 + \alpha^8 \right) }$ \\
$2/8$ &10010000 & $\alpha^{8}\,{{\frac{1 - \alpha^2 + \alpha^4}{1 + \alpha^4 + \alpha^8 + \alpha^{12}}}}$ & $\frac{{\left( 1 - \alpha^2 \right) }^2\,\left( 3 - 2\,\alpha^2 + 4\,\alpha^4 - 2\,\alpha^6 + 3\,\alpha^8 \right) }{16\,\left( 1 + \alpha^4 + \alpha^8 + \alpha^{12} \right) }$ \\
$2/8$ &10001000$^\ast$ & $\alpha^6\,\frac{1- \alpha^2}{{{1 - \alpha^{8}}}}$ & $\frac{{\left( 1 - \alpha^2 \right) }^2\,\left( 3 + 4\,\alpha^2 + 3\,\alpha^4 \right) }{16\,{\left( 1 + \alpha^2 \right) }^2\,\left( 1 + \alpha^4 \right) }$ \\
\noalign{\vskip 2pt}
\hline
\noalign{\vskip 2pt}
$3/8$ &11100000 & $\alpha^{10}\,{{\frac{1 - \alpha^6}{1 - \alpha^{16}}}}$ & $\frac{{\left( 1 - \alpha^2 \right) }^2\,\left( 15 + 44\,\alpha^2 + 71\,\alpha^4 + 80\,\alpha^6 + 71\,\alpha^8 + 44\,\alpha^{10} + 15\,\alpha^{12} \right) }{64\,{\left( 1 + \alpha^2 \right) }^2\,\left( 1 + \alpha^4 + \alpha^8 + \alpha^{12} \right) }$ \\
$3/8$ &10110000 & $\alpha^{8}\,{{\frac{1 - \alpha^4 + \alpha^6 - \alpha^8}{1 - \alpha^{16}}}}$ & $\frac{{\left( 1 - \alpha^2 \right) }^2\,\left( 15 + 28\,\alpha^2 + 39\,\alpha^4 + 48\,\alpha^6 + 39\,\alpha^8 + 28\,\alpha^{10} + 15\,\alpha^{12} \right) }{64\,{\left( 1 + \alpha^2 \right) }^2\,\left( 1 + \alpha^4 + \alpha^8 + \alpha^{12} \right) }$ \\
$3/8$ &10011000 & $\alpha^6\,{{\frac{1 - \alpha^4 + \alpha^8 - \alpha^{10}}{1 - \alpha^{16}}}}$ & $\frac{{\left( 1 - \alpha^2 \right) }^2\,\left( 15 + 28\,\alpha^2 + 23\,\alpha^4 + 16\,\alpha^6 + 23\,\alpha^8 + 28\,\alpha^{10} + 15\,\alpha^{12} \right) }{64\,{\left( 1 + \alpha^2 \right) }^2\,\left( 1 + \alpha^4 + \alpha^8 + \alpha^{12} \right) }$ \\
$3/8$ &11010000 & $\alpha^8\,{{\frac{1 - \alpha^2 + \alpha^4 - \alpha^8}{1 - \alpha^{16}}}}$ & $\frac{{\left( 1 - \alpha^2 \right) }^2\,\left( 15 + 28\,\alpha^2 + 39\,\alpha^4 + 48\,\alpha^6 + 39\,\alpha^8 + 28\,\alpha^{10} + 15\,\alpha^{12} \right) }{64\,{\left( 1 + \alpha^2 \right) }^2\,\left( 1 + \alpha^4 + \alpha^8 + \alpha^{12} \right) }$ \\
$3/8$ &10101000 & $\alpha^6\,{{\frac{(1 - \alpha^2)( 1 + \alpha^4 + \alpha^8) }{1 - \alpha^{16}}}}$ & $\frac{{\left( 1 - \alpha^2 \right) }^2\,\left( 15 + 12\,\alpha^2 + 23\,\alpha^4 + 16\,\alpha^6 + 23\,\alpha^8 + 12\,\alpha^{10} + 15\,\alpha^{12} \right) }{64\,{\left( 1 + \alpha^2 \right) }^2\,\left( 1 + \alpha^4 + \alpha^8 + \alpha^{12} \right) }$ \\
$3/8$ &10010100$^\ast$ & $\alpha^4\,{{\frac{(1 - \alpha^2)(1 + \alpha^4 + \alpha^{10})}{1 - \alpha^{16}}}}$ & $\frac{{\left( 1 - \alpha^2 \right) }^2\,\left( 15 + 12\,\alpha^2 + 7\,\alpha^4 + 16\,\alpha^6 + 7\,\alpha^8 + 12\,\alpha^{10} + 15\,\alpha^{12} \right) }{64\,{\left( 1 + \alpha^2 \right) }^2\,\left( 1 + \alpha^4 + \alpha^8 + \alpha^{12} \right) }$ \\
$3/8$ &11001000 & $\alpha^6\,{{\frac{1 - \alpha^2 + \alpha^6 - \alpha^{10}}{1 - \alpha^{16}}}}$ & $\frac{{\left( 1 - \alpha^2 \right) }^2\,\left( 15 + 28\,\alpha^2 + 23\,\alpha^4 + 16\,\alpha^6 + 23\,\alpha^8 + 28\,\alpha^{10} + 15\,\alpha^{12} \right) }{64\,{\left( 1 + \alpha^2 \right) }^2\,\left( 1 + \alpha^4 + \alpha^8 + \alpha^{12} \right) }$ \\
\noalign{\vskip 2pt}
\hline
\noalign{\vskip 2pt}
$2/9$ &110000000 &        $\alpha^{14}\,{{\frac{1 - \alpha^4}{1 - \alpha^{18}}}}$ &  $\frac{2\,{\left( 1 - \alpha^2 \right) }^2\,\left( 7 + 12\,\alpha^2 + 15\,\alpha^4 + 16\,\alpha^6 + 15\,\alpha^8 + 12\,\alpha^{10} + 7\,\alpha^{12} \right) }{81\,\left( 1 + \alpha^2 + \alpha^4 \right) \,\left( 1 + \alpha^6 + \alpha^{12} \right) }$ \\
$2/9$ &101000000 &        $\alpha^{12}\,{{\frac{1 - \alpha^2 + \alpha^4 - \alpha^6}{1 - \alpha^{18}}}}$ &  $\frac{2\,{\left( 1 - \alpha^2 \right) }^2\,\left( 7 + 3\,\alpha^2 + 15\,\alpha^4 + 7\,\alpha^6 + 15\,\alpha^8 + 3\,\alpha^{10} + 7\,\alpha^{12} \right) }{81\,\left( 1 + \alpha^2 + \alpha^4 + \alpha^6 + \alpha^8 + \alpha^{10} + \alpha^{12} + \alpha^{14} + \alpha^{16} \right) }$ \\
$2/9$ &100100000 &        $\alpha^{10}\,{{\frac{1 - \alpha^2 + \alpha^6 - \alpha^8}{1 - \alpha^{18}}}}$ & $\frac{2\,{\left( 1 - \alpha^2 \right) }^2\,\left( 7 + 3\,\alpha^2 + 6\,\alpha^4 + 7\,\alpha^6 + 6\,\alpha^8 + 3\,\alpha^{10} + 7\,\alpha^{12} \right) }{81\,\left( 1 + \alpha^2 + \alpha^4 + \alpha^6 + \alpha^8 + \alpha^{10} + \alpha^{12} + \alpha^{14} + \alpha^{16} \right) }$ \\
$2/9$ &100010000$^\ast$ & $\alpha^8\,{{\frac{1 - \alpha^2 + \alpha^8 - \alpha^{10}}{1 - \alpha^{18}}}}$ & $\frac{2\,{\left( 1 - \alpha^2 \right) }^2\,\left( 7 + 3\,\alpha^2 + 6\,\alpha^4 - 2\,\alpha^6 + 6\,\alpha^8 + 3\,\alpha^{10} + 7\,\alpha^{12} \right) }{81\,\left( 1 + \alpha^2 + \alpha^4 + \alpha^6 + \alpha^8 + \alpha^{10} + \alpha^{12} + \alpha^{14} + \alpha^{16} \right) }$ \\
\end{tabular}
\end{ruledtabular}
\label{table1}
\end{center}
\end{table*}

So far, we did not consider the cost of going from $x_{2,n+N-1}$ to $x_{2,n}$.
Denoting $z=x_{2,n+N-1}$ to simplify the expressions,
the new internal states after a $\Theta_{n+N}=0$ or $\Theta_{n+N}=1$ event are
\begin{eqnarray}
{\bf X}_0&=&(\sqrt{1-\alpha^2 z^2},\alpha z)
,\nonumber \\
{\bf X}_1&=&(\alpha\sqrt{1-z^2},\sqrt{1-\alpha^2+\alpha^2z^2})
,
\label{CRIT1}
\end{eqnarray}
respectively.
According to the general rules, the \DLM\ determines $\Theta_{n+N}$
by comparing the costs
\begin{eqnarray}
C_0&=&-\sqrt{1-\alpha^2z^2}\cos\theta -\alpha z\sin\theta
,\nonumber \\
C_1&=&-\alpha\sqrt{1-z^2}\cos\theta -\sqrt{1-\alpha^2+\alpha^2z^2}\sin\theta
,
\label{CRIT2}
\end{eqnarray}
for the two alternative internal states of Eq.~(\ref{CRIT1}).
The \DLM\ generates a $\Theta_{n+N}=1$ event if $C_1<C_0$.
After some rearrangements we obtain
\begin{eqnarray}
\tan\theta>\frac{\alpha z+\sqrt{1-\alpha^2+\alpha^2z^2}}
{\sqrt{1-\alpha^2z^2}+\alpha\sqrt{1-z^2}}
.
\label{CRIT3}
\end{eqnarray}
In general, $z$ is a function of $\alpha$.
Therefore, for a fixed $\alpha$, Eq.~(\ref{CRIT3}) sets an upperbound
to the input angle for which the \DLM\ can generate a periodic sequence.

As an illustration, we consider the sequence $\{00\ldots001\}$
in which there are $K$ 0 events and one 1 event.
The initial point for the periodic continuation
$\{00\ldots00100\ldots001,\ldots\}$ of this sequence is
given by Eq.~(\ref{CIRC6b}).
Let us assume that the \DLM\ starts from this initial point and
generates $K$ zero's, changing its internal state
from $\hat x^2$ to $z^2=\alpha^{2K}(1-\alpha^2)/(1-\alpha^{2K+2})$.
The \DLM\ will generate a $\Theta_{K+1}=1$ event if
\begin{eqnarray}
f(\alpha,K)\equiv\frac{1}{\sqrt{K}}-\frac{\alpha z+\sqrt{1-\alpha^2+\alpha^2z^2}}
{\sqrt{1-\alpha^2z^2}+\alpha\sqrt{1-z^2}} >0
.
\label{CRIT4}
\end{eqnarray}
In Fig.~\ref{alpha} we plot $f(\alpha,K)$ as a function of $\alpha$
for $K=57$ and $K=80$ (plots for other values of $K$ show the same
behavior).
From Fig.~\ref{alpha} and Eq.~(\ref{CRIT4}),
we conclude that the \DLM\ will indeed repeat the sequence $\{00\ldots001\}$
with $K=57$ ($K=80$) if $0.9967<\alpha<1$ ($0.9983<\alpha<1$).
Otherwise, if $\alpha$ is not within this range, the \DLM\
generates at least one extra 0 event and the \DLM\
does not return to the initial state $\hat x^2$.
Thus, if $\alpha$ is such that $f(\alpha,K)<0$,
the \DLM\ cannot generate the sequence
that gives an exact representation of $1/K$
(although it still gives an accurate approximation).

\begin{widetext}
\subsubsection{Variance of the periodic sequences}{\label{variance}}

Next, we compute the variance of the periodic, stationary state
$\{\hat x_{2,j+1}^2,\hat x_{2,j+2}^2,\ldots,\hat x_{2,j+N}^2\}$.
The expression of the variance reads

\begin{eqnarray}
\Delta^2&\equiv&
\frac{1}{N}\sum_{j=0}^{N-1}
\hat x_{2,n+j}^4
-\left(
\frac{1}{N}\sum_{j=0}^{N-1}
\hat x_{2,n+j}^2
\right)^2
.
\label{FIX2b}
\end{eqnarray}
Using
\begin{eqnarray}
x_{2,n+j+1}^4&=&
\alpha^4 x_{2,n+j}^4 + (1-\alpha^2)^2\Theta_{n+j+1}
+\alpha^2 (1-\alpha^2)^2\Theta_{n+j+1}x_{2,n+j}^2
,
\label{FIX2}
\end{eqnarray}
we obtain
\begin{eqnarray}
\frac{1}{N}\sum_{j=0}^{N-1}
\hat x_{2,n+j}^4
&=&
\frac{(1-\alpha^{2})^2}{1-\alpha^{4}}
\left[
{\bar \Theta}
+
\frac{2\alpha^{2}}{1-\alpha^{2N}}
\frac{1}{N}
\sum_{j=0}^{N-1}
\sum_{i=0}^{N-1}
\alpha^{2i}\Theta_{n+j+1} \Theta_{n+j-i}
\right]
,\nonumber \\
&=&
\frac{1-\alpha^{2}}{1+\alpha^{2}}
\left[
{\bar \Theta}
+
\frac{2\alpha^{2}}{1-\alpha^{2N}}
\frac{1}{N}
\sum_{j=0}^{N-1}
\sum_{i=0}^{N-1}
\alpha^{2i}\Theta_{n+j+i+1} \Theta_{n+j}
\right]
,\nonumber \\
&=&
\frac{1-\alpha^{2}}{1+\alpha^{2}}
\left[
\frac{1+\alpha^{2N}}{1-\alpha^{2N}}
{\bar \Theta}
+
\frac{2\alpha^{2}}{1-\alpha^{2N}}
\frac{1}{N}
\sum_{j=0}^{N-1}
\sum_{i=0}^{N-2}
\alpha^{2i}\Theta_{n+j+i+1} \Theta_{n+j}
\right]
,
\label{FIX3a}
\end{eqnarray}
where in the last step, we have taken out from the double sum,
all terms of the form $\Theta_{n+j} \Theta_{n+j}$.
\medskip
\end{widetext}

In Table~\ref{table1} we present analytical results for the
initial points and variances
of some simple sequences $\{\Theta_1,\ldots,\Theta_q\}$
with periods $q$ and $\bar\Theta=p/q$ where both $p$ and $q$ are integers.
If $p$ and $q$ have a common factor $c$, as is the case for $p=2$ and $q=8$,
the problem simplifies to the case $(p/c)/(q/c)$.
The sequences $\{\Theta_1,\ldots,\Theta_q\}$
marked with a $^\ast$ yield the smallest variance $\Delta^2$.
These are exactly the sequences that the \DLM\ generates
in the stationary regime, provided $1-\alpha$ is sufficiently
small ($\alpha=0.99$ is sufficient for the $(p,q)$-cases
presented in Table~\ref{table1}).

In general, any sequence of 0's and 1's that begins with a 1,
can be viewed as a concatenation of
subsequences that start with a 1 followed by one or more 0's.
The examples in Table~\ref{table1} suggest that
the sequences $\{\Theta_1,\ldots,\Theta_q\}$ with the smallest variance
consist of $n_1$ subsequences of length $L_1\equiv\lfloor q/p \rfloor\le q/p$
and $n_2$ subsequences of length $L_2\equiv\lceil q/p\rceil = L_1+1 > q/p$.
We have not been able to prove that in general
this is the structure of the minimum variance solution.
However, the relation between the minimum variance solution
and the ground state configurations of a one-dimensional lattice
model to be discussed next, suggests that this may well be the case.

\subsubsection{Generalized one-dimensional Wigner lattice}{\label{wigner}}

From Eqs.~(\ref{FIX2b}) and (\ref{FIX3a}), it follows
that minimizing the variance $\Delta^2$
is tantamount to finding the periodic sequence
$\{\Theta_{n+1},\Theta_{n+2},\ldots,\Theta_{n+N}; \Theta_{n}=\Theta_{n+N}\}$
that minimizes the last term in Eq.~(\ref{FIX3a}), subject
to the constraint $\bar \Theta= N^{-1}\sum_{j=1}^{N}\Theta_{n+j}$.
We now show that solving for the sequence
that yields the lowest variance amounts to finding
the ground state configuration of a classical many-body
system.

If we interpret $\Theta_{j}=0$ (1) as the absence (presence) of a particle
at the site $j$ of a one-dimensional lattice,
then the last term in Eq.~(\ref{FIX3a}) can be written as
\begin{eqnarray}
H&=&
\sum_{j=0}^{N-1}
\sum_{k=0}^{N-2}
V_k n_j n_{j+k}
,
\label{FIX4}
\end{eqnarray}
where $V_k=\alpha^{2k}$ and $n_i$ takes the value
0 or 1 if the site $i$ is empty or occupied, respectively.
Clearly, if $\alpha<1$,
the potential $V_k$ satisfies the two conditions
\begin{eqnarray}
\lim_{k\rightarrow\infty} V_k=0
\quad,\quad 
V_{k-1}+V_{k+1}\ge 2V_k\quad,\quad k>1
.
\label{FIX5}
\end{eqnarray}
The density of particles
$\rho\equiv N^{-1}\sum_{j=1}^{N} n_j$
is given by $\rho=\bar \Theta$.

In the limit $N\rightarrow\infty$, the problem of finding the ground
state configuration of particles for a system defined by the Hamiltonian
\begin{eqnarray}
H&=& \sum_{i\not=j} V_{|i-j|} n_i n_j
,
\label{FIX6}
\end{eqnarray}
and satisfying the two conditions Eq.~(\ref{FIX5})
was solved by Hubbard~\cite{HUBB78}.
For any density $\rho=p/q$ where $p$ and
$q$ are integers with no common factor,
the ground state of Eq.~(\ref{FIX6}) is periodic
with period $q$ and $p$ particles in each period~\cite{HUBB78}.

Hubbard gives an algorithm to generate the
ground state configuration for a pair ($p$,$q$)
and calls these ground state configurations
generalized one-dimensional Wigner lattices~\cite{HUBB78}.
His algorithm also generates the sequences $\{\Theta_1,\ldots,\Theta_q\}$
in Table~\ref{table1} that are marked with a $^\ast$.
This is not a surprise: The periodic sequences
with the smallest variance $\Delta^2$ are also
the ground state configurations of model Eq.~(\ref{FIX6}).
Extensive numerical tests for $q=2,\ldots,10000$ and $1\le p<q$ (results not shown) confirm
that the ground state configurations generated by
Hubbard's algorithm are the same as the
periodic sequences generated by the \DLM\ in the stationary regime,
for a fixed input ${\bf y}=(y_1,y_2)$, $y_1^2=q$, $y_2^2=p$,
and sufficiently small $\alpha$.

\begin{figure}[t]
\begin{center}
\includegraphics[width=8.65cm]{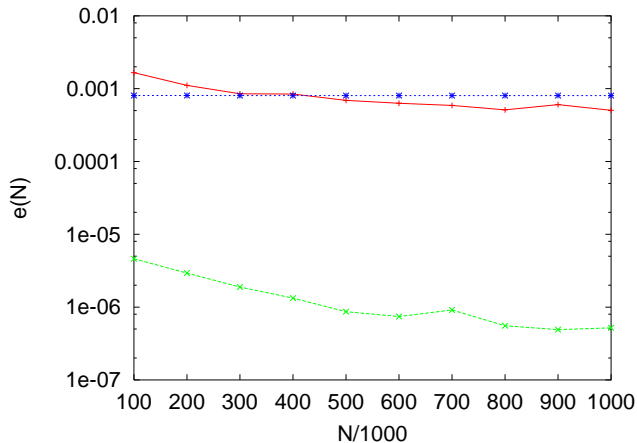}
\caption{(color online)
The error $e(N)$ defined by Eq.~(\ref{PER0})
as a function of the number of events $N$ for
$M=100$ (corresponding to 101 different input angles).
For each $m=0,\ldots,M$, we generate $N$ input events,
each input event carrying the message
${\bf y}_{n+1}=(\cos(\arcsin\sqrt{m/M}),\sin(\arcsin\sqrt{m/M}))$.
Note that ${\bf y}_{n+1}$ is a vector of rational numbers.
Solid (red) line: Probabilistic Bernoulli-type processor (see Section~\ref{sec2});
Dashed (green) line: Deterministic learning machine (see Section~\ref{sec3}).
Dotted (blue) line: Modified
deterministic learning machine, see Eq.~(\ref{PER1}).
In all the \DLM\ simulations, $\alpha=0.9995$
and the first 10000 event were discarded to
allow the \DLM\ to approach the stationary state.
}
\label{performance1}
\end{center}
\end{figure}

\begin{figure}[t]
\begin{center}
\includegraphics[width=8.65cm]{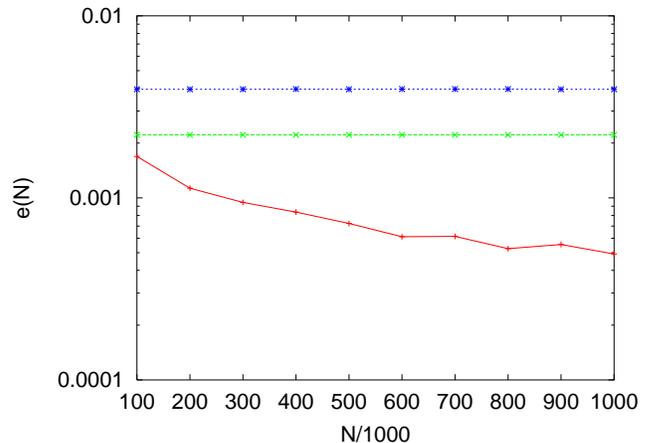}
\caption{(color online)
Same as in Fig.~\ref{performance1} except
that the input events carry the message
${\bf y}_{n+1}=(\cos(m\pi/2M),\sin(m\pi/2M))$
for $m=0,\ldots,M$.
}
\label{performance2}
\end{center}
\end{figure}

\subsection{Performance analysis}{\label{sec4}}

The non-analytic character of the \DLM\ algorithm
and the complicated dependence on the parameter $\alpha$
make it difficult for us to proof more rigorous results
about the \DLM\ dynamics than those presented earlier.
On the other hand, it is very easy to study the dynamics numerically.
Extensive simulation work (results not shown) demonstrate that, with a proper choice
of $\alpha$ (see Sections~\ref{resol} and~\ref{lower}),
a \DLM\ can encode all rational numbers $n/N$ for $n=0,\ldots,N$.
Thus, for each input angle $\psi$ for which $\sin^2(\psi-\phi)$
is a rational number, there is the stationary state in which the \DLM\
generates to a unique, periodic sequence (with minimum variance)
of 1's and 0's from which the value of $\sin^2(\psi-\phi)$
can be determined with very high precision.

The update rule that the \DLM\ uses is quite subtle,
and we demonstrate this by changing the rules Eqs.~(\ref{CIRC3}) and (\ref{CIRC4})
to
\begin{eqnarray}
x_{2,n+1}^2&=&\alpha^2 x_{2,n}^2 + (1-\alpha^2)\widehat\Theta_{n+1},
\nonumber \\
\widehat\Theta_{n+1}&=&
\frac{1}{2}\left(1-\frac{x_{2,n}^2-y_{2,n}^2}{|x_{2,n}^2-y_{2,n}^2|}\right)
.
\label{PER1}
\end{eqnarray}
For $0\le\theta\le\pi/2$ (the case of interest for the present analysis),
this rule tells the machine to rotate its internal
vector towards the input vector ${\bf y}_{n+1}=(y_{1,n+1},y_{2,n+1})$.
In contrast, the \DLM\ that operates according to the
rules of Eqs.~(\ref{CIRC3}) and (\ref{CIRC4})
may decide to rotate its internal
vector away from the input vector.

For a quantitative comparison of the
performance of the probabilistic processor,
the \DLM\ defined by the rules of Eqs.~(\ref{CIRC3}) and (\ref{CIRC4}),
and the \DLM\ defined by the rules of Eq.~(\ref{PER1}),
we carry out the procedure that follows:
\begin{enumerate}
\item set $M=100$ and choose $\phi\in[0,360[$ randomly
\item for each $m=0,\ldots,M$
\item set $e_m(N)=0$
\item compute $\psi_m-\phi=\arcsin\sqrt{m/M}$
\item for $n=1,\ldots,N$
\item generate an input event carrying the message
${\bf y}_{n}=(\cos(\arcsin\sqrt{m/M}),\sin(\arcsin\sqrt{m/M}))$
\item count the number $K$ of 1 events generated by the processor
\item end loop over $n$
\item compute $\psi_m^\prime-\phi=\arcsin\sqrt{K/N}$
\item set $e_m(N)=e_m(N)+(\psi_m-\psi_m^\prime)^2$
\item end loop over $m$
\end{enumerate}
The number accumulated in $e_m(N)$ yields
\begin{eqnarray}
e(N)&=&
\sqrt{ \frac{1}{M+1}\sum_{m=0}^M e_m(N) }
\nonumber \\
&=&
\sqrt{\frac{1}{M+1}\sum_{m=0}^M (\psi_m-\psi_m^\prime)^2}
,
\label{PER0}
\end{eqnarray}
which is the error averaged over $M+1$ different pairs
of (input,output) angles for a fixed number $N$ of input messages.
Fig.~\ref{performance1} shows the error $e(N)$
as a function of the number of events $N$.
In this case, $\cos^2(\psi_m-\phi)$ and $\sin^2(\psi_m-\phi)$ are rational
numbers and the results of Fig.~\ref{performance1} confirm that
the \DLM\ performs very well,
much better than the probabilistic processor.
Fig.~\ref{performance1} also shows that
replacing the update rule Eq.~(\ref{CIRC5})
by Eq.~(\ref{PER1}) has a large impact on the performance
of a deterministic learning machine.

If we replace $\psi_m-\phi=\arcsin\sqrt{m/M}$ by
$\psi_m-\phi=m\pi/2M$ in step 4 of the procedure described earlier,
then $\sin^2(\psi_m-\phi)$ is not necessarily
a rational number and this affects the performance
of the \DLM, as shown in Fig.~\ref{performance2}.
A closer look at the \DLM\ data for different $m$ (results
not shown) reveals that the large increase of the error
is due to the relatively poor accuracy for $m\approx0$
and $m\approx M$.
This is hardly a surprise:
From Sections~\ref{resol} and~\ref{lower} we know
that the choice of $\alpha$ is more important
for $\theta\approx n\pi/2$ than it is
for $\theta\approx n\pi/2+\pi/4$.
Therefore, if optimal performance for all $\theta$
is crucial, it is necessary to adjust $\alpha$ dynamically
by another learning process. We leave this topic
for future research.

\begin{figure*}[t]
\begin{center}
\includegraphics[width=16cm]{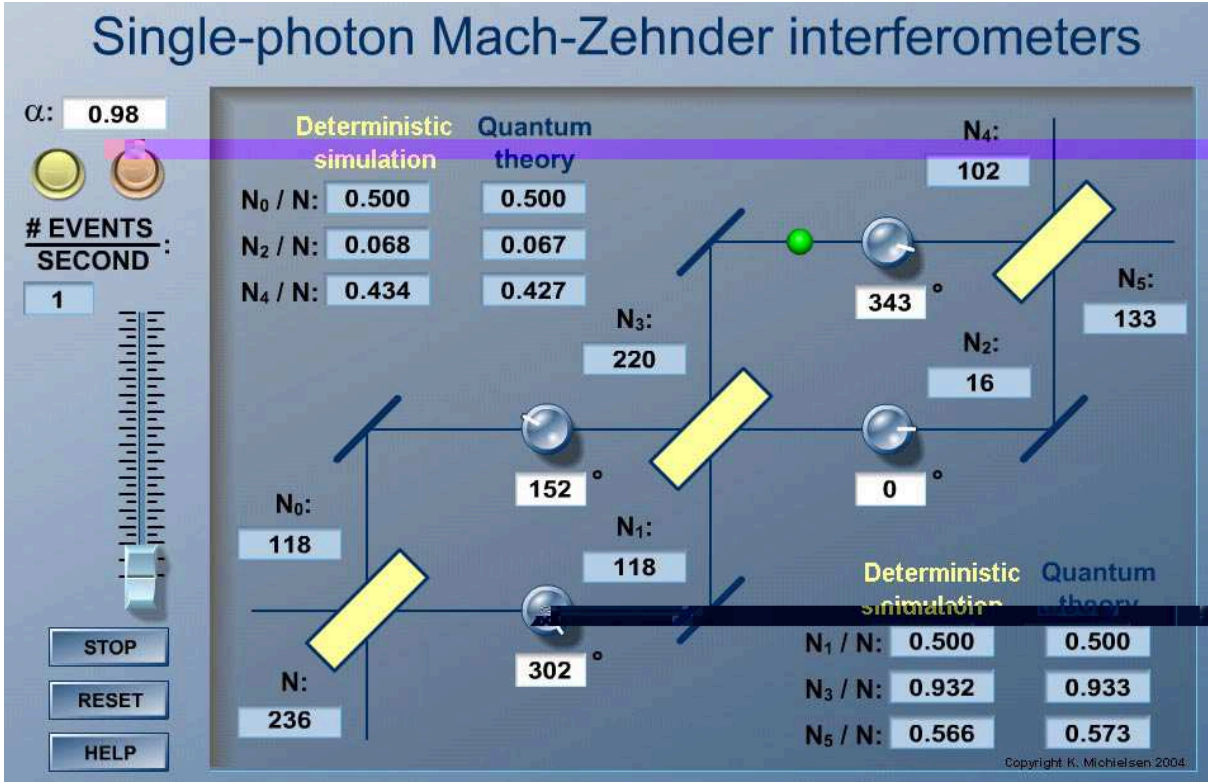}
\caption{(color online)
Snapshot of the event-by-event simulator
of two concatenated Mach-Zehnder interferometers~\cite{MZIdemo}.
The main panel shows the layout of the interferometer.
Particles emerge from a source (not shown)
located at the bottom of the left-most vertical line.
After leaving the first beam splitter in either
the vertical or horizontal direction,
the particles experience time delays
that are specified by the controls on the lines.
In this example, the time delays correspond
to the phase shifts
$\phi_0=152^\circ$,
$\phi_1=302^\circ$,
$\phi_2=0^\circ$,
and $\phi_3=342^\circ$ in the quantum mechanical description.
The thin, $45^\circ$-tilted lines act as perfect mirrors.
When a particle leaves the system at the top right,
it adds to the count of either detector $N_4$ or $N_5$.
Additional detectors ($N_0$, $N_1$, $N_2$, and $N_3$)
count the number of particles on the corresponding lines.
The other cells give the ratio of the detector counts
to the total number of particles (messages) processed
and also the corresponding probability of
the quantum mechanical description.
At any time, the user can choose between a
strictly deterministic and a probabilistic event-by-event
simulation by pressing the buttons at the top of the control panel.
}
\label{twomzi}
\end{center}
\end{figure*}

\section{Relevance of the learning process}{\label{sec5}}

The fundamental difference between the simple
probabilistic processor of Section~\ref{sec2}
and the \DLM\ of Section~\ref{sec3} is that
the latter has a learning capability.
Elsewhere, we have shown that learning is an essential ingredient
of networks of probabilistic or deterministic processors
that are able to simulate, event-by-event,
quantum interference phenomena and
universal quantum computation~\cite{KRAED05,HRAED05a,MICH05,HRAED05b}.
The fundamental reason for this is that some form of communication
between individual events is required in order to
simulate (classical or quantum) interference phenomena.
Although the Bernoulli-type probabilistic processor of Section~\ref{sec2}
satisfies our criteria 1 and 2 of Section~\ref{sec1} for an efficient processor, it generates
uncorrelated events and any form of communication between events is absent.
Therefore the probabilistic processor of Section~\ref{sec2}
cannot simulate interference phenomena
but the \DLM\ of Section~\ref{sec3} can because it has
the additional feature of being able to learn from previous events.

As a non-trivial illustration of the importance
of the learning process, we consider the
interferometer depicted in Fig.~\ref{twomzi}~\cite{MZIdemo}.
This interferometer consists of two chained Mach-Zehnder interferometers~\cite{BORN64}.
Photons leave the source (not shown)
located at the bottom of the left-most vertical line.
The beam splitters, represented by the large rectangles,
transmit or reflect photons with probability 1/2.
After leaving the first beam splitter in
the vertical or horizontal direction,
the photons experience a time delay
that is determined by the length of the optical path
from one beam splitter to the next.
The length of each path is variable,
as indicated schematically by the controls on the horizontal lines.
In a wave mechanical description, the time delays correspond
to changes in the phase of the wave.
The thin, $45^\circ$-tilted lines act as perfect mirrors.

In quantum theory, the presence of photons in
the input modes 0 or 1 of the interferometer is represented
by the probability amplitudes ($a_0,a_1)$~\cite{BAYM74}.
According to quantum theory, the amplitudes ($b_0,b_1)$
of the photons in the output modes 0 and 1 of
a beam splitter are given by~\cite{BAYM74}
\begin{eqnarray}
\left(
\begin{array}{c}
b_0\\
b_1
\end{array}
\right)
=
\frac{1}{\sqrt{2}}
\left(
\begin{array}{cc}
1&i\\
i&1
\end{array}
\right)
\left(
\begin{array}{c}
a_0\\
a_1
\end{array}
\right).
\label{BS0}
\end{eqnarray}
The amplitudes to observe a
photon in the output modes 0 and 1 of
one Mach-Zehnder interferometer of Fig.~\ref{twomzi}
are given by
\begin{eqnarray}
\left(
\begin{array}{c}
b_2\\
b_3
\end{array}
\right)
=
\left(
\begin{array}{cc}
1&i\\
i&1
\end{array}
\right)
\left(
\begin{array}{cc}
e^{i\phi_0}&0\\
0&e^{i\phi_1}
\end{array}
\right)
\left(
\begin{array}{c}
b_0\\
b_1
\end{array}
\right)
.
\label{MZ1}
\end{eqnarray}
The amplitudes to observe a
photon in the output modes 0 and 1 of
two chained Mach-Zehnder interferometers are given by
\begin{eqnarray}
\left(
\begin{array}{c}
b_4\\
b_5
\end{array}
\right)
=
\left(
\begin{array}{cc}
1&i\\
i&1
\end{array}
\right)
\left(
\begin{array}{cc}
e^{i\phi_2}&0\\
0&e^{i\phi_3}
\end{array}
\right)
\left(
\begin{array}{c}
b_2\\
b_3
\end{array}
\right)
.
\label{MZ2}
\end{eqnarray}
In Eqs.~(\ref{MZ1}) and (\ref{MZ2}), the entries
$e^{i\phi_j}$ for $j=0,1,2,3$ implement the phase shifts that result
from the time delays on the corresponding path
(including the phase shifts due to the presence of the perfect mirrors).
The probability to detect a photon in either output mode 0 or 1
of the two chained Mach-Zehnder interferometers are given by
$|b_4|^2$ or $|b_5|^2$, respectively.

Using \DLM\ networks, it is possible to reproduce the wave-like behavior
by an event-by-event, particle-like, simulation without
using wave mechanics~\cite{KRAED05,HRAED05a,MICH05,HRAED05b}.
Elsewhere~\cite{KRAED05,HRAED05a,MICH05,HRAED05b} we have shown
that \DLM\ networks can simulate, event by event,
single-photon beam splitter and (modified) Mach-Zehnder
interferometer experiments~\cite{GRAN86,BRAI03}.

Fig.~\ref{twomzi} shows
the schematic diagram of the \DLM\ network
that performs the event-by-event simulation of the
two chained Mach-Zehnder interferometers~\cite{MZIdemo}.
Particles emerge one-by-one from a source (not shown)
located at the bottom of the left-most vertical line.
At any time, there is at most one particle
(represented by the small sphere) in the system.
The number of particles that have left the source is given by $N$.

Each particle carries its own clock.
There is a one-to-one correspondence
between the direction of the hand of the clock
and the message ${\bf y}_{n+1}=(y_{1,n+1},y_{2,n+1})$.
The clock is read and manipulated by the beam splitters,
represented by the large rectangles.
Each beam splitter contains two \DLMS~\cite{KRAED05}.
The internal structure of the \DLM\ network
that performs the task of a beam splitter is described in detail
elsewhere~\cite{KRAED05,HRAED05a,MICH05,HRAED05b},
so there is no need to repeat it here.
Of course, these networks are the same for the three beam splitters.

After leaving the first beam splitter in either
the vertical or horizontal direction (but never in both),
the particle experiences a time delay
that is determined by the controls on the lines.
This time delay is implemented as a rotation
of the hand of the particle's clock.
When a particle leaves the system at the top right,
it adds to the count of either detector $N_4$ or $N_5$.
Additional detectors ($N_0$, $N_1$, $N_2$, and $N_3$)
count the number of particles on the corresponding lines.
The label of $\phi_j$ in the quantum mechanical
description is the same as the label of the corresponding counter $N_j$.
The other cells give the ratio of the detector counts
to the total number of particles (messages) processed
and also the corresponding probability of
the quantum mechanical description.
At any time, the user can choose between a
strictly deterministic and a probabilistic event-by-event
simulation by pressing the buttons at the top of the control panel.
We emphasize that this \DLM-based simulation is dynamic and adaptive
in all respects: During the simulation, the user
can change any of the controls and after a short
transient period, the \DLM-network generates output events according
to the quantum mechanical probabilities.

The snapshot in Fig.~\ref{twomzi} is taken after
$N=236$ particles have been generated by the source
(with one particle still under way).
The numbers in the various corresponding fields
clearly show that even after a modest number of
events, this event-by-event simulation reproduces
the quantum mechanical probabilities.
Of course, this single snapshot is not a proof that
the event-by-event simulation also works for
other choices of the delays.

An event-by-event simulation correctly reproduces
the quantum mechanical probabilities
if and only if $N_j/N\approx |b_j|^2$ for $j=0,1,2,3$,
for any choice of the delays (phases) $\phi_j$.
Very extensive tests,
reported elsewhere~\cite{KRAED05,HRAED05a,MICH05,HRAED05b}
demonstrate that \DLM-networks accurately reproduce the probabilities
of the quantum theory.

In the event-by-event simulation, interference is a direct result
of the learning process that takes place in each \DLM.
In the case at hand, the three (identical) beam splitters
contain the learning machines.
We emphasize that there is no direct communication
between the different beam splitters.
All the information is carried by the particle
while it is routed through the network.
This is essential for the simulation to satisfy
the physical criterion of causality.

\section{Summary}{\label{sec6}}

In this paper we ask ourselves the question what the optimal design of a processor,
which can process and count incoming individual objects carrying information represented by an angle
$\psi$ but which cannot measure $\psi$ directly, would be if it has to give the most accurate estimate of the 
angle $\psi$. In other words, how can we simulate the operation of a photon polarizer?

First, we construct a processor operating according to the rules of probability theory. This
so-called probabilistic processor uses random numbers to transform the incoming angle $\psi$, that is
the information carried by the incoming single objects, into a sequence of discrete output events labeled 
by $\pm 1$. The numbers of $+1$ and $-1$ events only depend on the difference $\theta =\psi-\phi$ between
the unknown angle $\psi$ and the orientation $\phi$ of the processor.
We design the probabilistic processor such that the result of the transformation process is 
probabilistic (Bernoulli trials), rotational invariant
and satisfies the criteria 1 and 2 of Section~\ref{sec1}.
For fixed $\phi$ and $N$ incoming
objects, the observer, using the probabilistic processor to measure $\psi$ as accurate as possible, will get
most out of the data if the processor sends $N\cos ^2 \theta$ ($N\sin ^2 \theta$) events to the apparatus
that detects the $+1$ ($-1$) events. The number of angles $\theta$ that the observer can distinguish is proportional
to $\sqrt N$. The probabilistic processor is thus a model for an ideal polarizer. It produces data that agrees with Malus ' law.
However, it is important to note that to obtain this result we do not use any law of physics in the design of the processor. 
We do not use the probability distributions derived in quantum theory to generate the $\pm 1$ events but we
design the probabilistic processor in such a way that these probability distributions come out as a result of efficient processing
of incoming data by the processor.
Hence, we can ask the following important question. Can also other quantum phenomena appear as a result of efficient data processing?

In order to answer this question we follow another route. Although the Bernoulli type probabilistic processor can
simulate the classical and quantum properties of a photon polarizer, it cannot simulate interference phenomena.
To overcome this problem we could design a probabilistic processor that does not generate Bernoulli events but
correlated output events. However, we  choose to design processors that use
a deterministic algorithm with a
primitive learning capability to transform the incoming events into a sequence of discrete output events. This type of
processors we call deterministic processors or deterministic learning machines.

Therefore, as a second step, we construct a deterministic processor that models a photon polarizer, that is a deterministic processor that
generates output events according to Malus' law.
Just as the probabilistic processor, the deterministic processor has one input channel and two output channels labeled by
$+1$ and $-1$, respectively.
Apart from this the deterministic processor also has an internal vector with two real entries.
The input messages to the deterministic processor are unit vectors ${\bf y} _{n+1}=(\cos\theta _{n+1}, \sin\theta_{n+1})$
for $n=0,\ldots ,N$ and where $\theta_{n}=\psi_n-\phi$.
The deterministic processor learns from the input events by updating its internal vector and uses this internal vector in a
completely deterministic decision process to send out a $+1$ or a $-1$ event. Hence, the order in which the $+1$ and $-1$ events are sent 
is deterministic.
Apart from being deterministic, the result of the transformation process
is rotational invariant and satisfies criteria 1 and 2 of Section~\ref{sec1},
which are exactly the same requirements as the ones
used to construct the probabilistic processor.
Further analysis of the output events shows that the number of $+1$ and $-1$ output events agrees with Malus' law.
Hence, the photon polarizer can also be modelled by a deterministic processor. As in the case of the probabilistic processor,
also in this case we did not use any laws of physics to design the processor.
The number of angles $\theta$ that the observer, using the deterministic processor to measure $\psi$,
can distinguish is equal to $N+1$. Hence, in this respect the deterministic processor performs much better than the probabilistic one.
However, the more important and fundamental difference between the probabilistic and the deterministic processor is that the latter has a learning
capability. Learning is an essential ingredient to simulate interference phenomena since it correlates the output events.
As an example we show the event-by-event simulation of photons routed through two chained Mach-Zehnder interferometers by using a network of deterministic processors. We show that the quantum mechanical probabilities are also reproduced for this interference experiment.

In conclusion, processors that efficiently process incoming data in the form of single events can simulate some quantum phenomema, such as the recovery of Malus' law for a photon polarizer.
However, in order to simulate quantum interference the processor
should in addition have the capability of learning.
Most importantly, the present work demonstrates that viewing quantum systems as efficient data
processors provides a framework to construct adaptive, dynamical systems that can
simulate quantum interference on an event-by-event basis, without using concepts of quantum theory.

\section*{Acknowledgment}
Support from the NAREGI Nanoscience Project, Ministry of Education,
Culture, Sports, Science and Technology, Japan is gratefully acknowledged.

\appendix
\section{On the use of the Cram\'er-Rao inequality}

In Frieden's approach the Cram\'er-Rao inequality
plays a central role to motivate the use of the Fisher information
as a measure of the expected error in measurements~\cite{FRIE99}.
From probability theory it is well known that the Cram\'er-Rao inequality
sets a lower bound to the variance of an estimator~\cite{TREE68,FRIE99,JAYN03}.
Here we prove that, within the limitations set by our design criteria,
the estimation procedure is {\sl efficient} in the
sense that it satisfies the Cram\'er-Rao inequality with equality~\cite{TREE68,FRIE99}
and that this inequality
reduces to a trivial identity that contains no information~\cite{JAYN03}.
For convenience of the reader, we repeat the derivation of the Cram\'er-Rao inequality
for the case of interest.

Writing Eq.({\ref{STOC2}}) as
\begin{equation}
\sum_{x=\pm1} (x-f(\theta)) p(x|\theta)=0
,
\label{APPE4}
\end{equation}
and taking the derivative with respect to $\theta$ we obtain
\begin{equation}
\sum_{x=\pm1} (x-f(\theta))\frac{\partial p(x|\theta)}{\partial\theta}
=\frac{\partial f(\theta)}{\partial\theta}
.
\label{APPE5}
\end{equation}
Rewriting Eq.({\ref{APPE5}}) as
\begin{widetext}
\begin{equation}
\sum_{x=\pm1} \left[(x-f(\theta))\sqrt{p(x|\theta)}\right]
\left[\frac{1}{\sqrt{p(x|\theta)}}\frac{\partial p(x|\theta)}{\partial\theta}\right]
=\frac{\partial f(\theta)}{\partial\theta}
,
\label{APPE6}
\end{equation}
and using the Schwartz inequality gives the Cram\'er-Rao inequality
\begin{equation}
\left\{\sum_{x=\pm1} (x-f(\theta))^2 p(x|\theta)\right\}
\left\{\sum_{x=\pm1} \frac{1}{p(x|\theta)}
\left[\frac{\partial p(x|\theta)}{\partial\theta}\right]^2\right\}
=
\hbox{Var}(x) I_F
\ge
\left(\frac{\partial f(\theta)}{\partial\theta}\right)^2
.
\label{APPE8}
\end{equation}
\end{widetext}
where
\begin{equation}
I_F=\sum_{x=\pm1} \frac{1}{p(x|\theta)}
\left[\frac{\partial p(x|\theta)}{\partial\theta}\right]^2
,
\label{APPE9}
\end{equation}
is the Fisher information~\cite{TREE68,FRIE99,BLAH91}.
With the use of Eq.({\ref{STOC1}}) we can write $I_F$ as
\begin{equation}
I_F=\frac{1}{p(1|\theta)(1-p(1|\theta))}
\left[\frac{\partial p(1|\theta)}{\partial\theta}\right]^2
,
\label{APPE10}
\end{equation}
which is identical to Eq.~(\ref{STOC9}).
Any estimation procedure that satisfies the bound in Eq.~(\ref{APPE8}) with
equality is called {\sl efficient}~\cite{TREE68,FRIE99}.
Using Eq.~(\ref{STOC2}) and
$\hbox{Var}(x)=\langle x^2\rangle-\langle x\rangle^2=
4p(1|\theta)(1-p(1|\theta))$
we find
\begin{eqnarray}
\hbox{Var}(x)I_F &=& 4
\left[\frac{\partial p(1|\theta)}{\partial\theta}\right]^2
=\left[\frac{\partial f(\theta)}{\partial\theta}\right]^2
.
\end{eqnarray}
Hence, the inequality Eq.~({\ref{APPE8}}) reduces to a trivial identity
from which we cannot deduce anything useful~\cite{JAYN03}.

\raggedright

\end{document}